\documentclass[notitlepage,english,aps,floats,onecolumn,showpacs,nofootinbib,floatfix]{revtex4-2}

\usepackage{pslatex}
\usepackage[T1]{fontenc}
\usepackage[latin1]{inputenc}
\usepackage[dvips]{graphicx}
\usepackage{epsfig}
\usepackage{longtable}
\usepackage{float}
\usepackage{calc}
\usepackage{ifthen}
\usepackage{amsmath}
\usepackage{amssymb}

\usepackage{ulem}

\usepackage{color}

\usepackage{xcolor}
\usepackage{tikz}

{
{
{
\newcommand{\bea}{\begin{eqnarray}}
\newcommand{\eea}{\end{eqnarray}}

\newcommand{\nc}{\newcommand}
\nc{\renc}{\renewcommand}
\nc{\eqs}[2]{\mbox{Eqs.~(\ref{#1},\,\ref{#2})}}
\nc{\eq}[1]{\mbox{Eq.~(\ref{#1})}}
\nc{\figs}[2]{\mbox{Figs.~(\ref{#1},\,\ref{#2})}}
\nc{\fig}[1]{\mbox{Fig~.(\ref{#1})}}
\nc{\be}[1]{\begin{equation} \mbox{$\label{#1}$}}
\nc{\ee}{\vspace{0.1cm}\end{equation}}

\newcommand{\bean}{\begin{eqnarray*}}
\newcommand{\eean}{\end{eqnarray*}}

\def\GeV{{\rm \ GeV}}

\def\lae{\;^{<}_{\sim} \;} \def\gae{\; ^{>}_{\sim} \;}

\begin{document}

\title{Q-balls in Non-Minimally Coupled Palatini Inflation and their Implications for Cosmology}

\author{A. K. Lloyd-Stubbs and J. McDonald}
\email{a.lloyd-stubbs@lancaster.ac.uk}
\email{j.mcdonald@lancaster.ac.uk}
\affiliation{Dept. of Physics,  
Lancaster University, Lancaster LA1 4YB, UK}

\begin{abstract}

We demonstrate the existence of Q-balls in non-minimally coupled inflation models with a complex inflaton in the Palatini formulation of gravity. We show that there exist Q-ball solutions which are compatible with inflation and we derive a window in the inflaton mass squared for which this is the case. In particular, we confirm the existence of Q-ball solutions with $\phi \sim 10^{17}-10^{18} \GeV $, consistent with the range of field values following the end of slow-roll Palatini inflation. We study the Q-balls and their properties both numerically and in an analytical approximation. The existence of such Q-balls suggests that the complex inflaton condensate can fragment into Q-balls, and that there may be an analogous process for the case of a real inflaton with fragmentation to neutral oscillons.  We discuss the possible post-inflationary cosmology following the formation of Q-balls, including an early Q-ball matter domination (eMD) period and the effects of this on the reheating dynamics of the model, gravitational wave signatures which may be detectable in future experiments, and the possibility that Q-balls could lead to the formation of primordial black holes (PBHs). In particular, we show that Palatini Q-balls with field strengths typical of inflaton condensate fragmentation can directly form black holes with masses around 500 kg or more when the self-coupling is $\lambda = 0.1$, resulting in very low (less than 100 GeV) reheating temperatures from black hole decay, with smaller black hole masses and larger reheating temperatures possible for smaller values of $\lambda$. Q-ball dark matter from non-minimally coupled Palatini inflation may also be a direction for future work.

\end{abstract}
 \pacs{}
 
\maketitle

\section{Introduction} 

Non-minimally coupled inflation models \cite{sbb}, \cite{bezrukov08} have been shown to be a promising framework for successful inflation, and can be considered in either the metric or Palatini formalisms of gravity. In this work we consider a complex scalar inflaton, $\Phi$, charged under a global $U\left(1 \right)$ symmetry, with a non-minimal coupling to gravity $2\xi \left|\Phi \right|^{2}$ in the Palatini formulation of gravity. In the Palatini formalism the spacetime metric, $g_{\mu \nu}$, and the spacetime connection, $\Gamma$, are defined independently of one another (see \cite{palatini}, \cite{ferraris82}) . In the standard metric formulation of general relativity, the spacetime connection takes the form of the Levi-Civita connection, and the Riemann tensor as a result has an underlying dependence on the spacetime metric. In the Palatini formalism this is not the case, and the only underlying dependence on the metric in the scalar curvature arises from the contraction of the Ricci tensor.

In inflation models with a non-minimal coupling (or other higher-order curvature terms), we can perform a conformal transformation on the metric of the theory in order to perform the inflationary calculations in a frame where the scalar is minimally coupled to gravity and therefore the gravitational dynamics are equivalent to general relativity (GR). This is more straightforward in a model in the Palatini formulation compared to the metric formulation, since there is no intrinsic dependence of the Ricci scalar on the metric and therefore only explicit factors of the spacetime metric need to be transformed. In the metric formulation, additional kinetic terms will arise due to the additional transformation of the Ricci scalar and can make for a more unwieldy calculation. In GR, the metric and Palatini formulations are equivalent, however one finds that the results of the inflationary calculations are different in non-minimally coupled  models and depend on the formalism used. Palatini inflation gives a much smaller value of the tensor-to-scalar ratio, $r$, than metric inflation, and generally requires a larger non-minimal coupling \cite{bauer08}. Palatini inflation can comfortably proceed with a sub-Planckian inflaton \cite{bauer08}, and can alleviate the issue of unitarity violation in non-minimally coupled Higgs inflation \cite{bauer11}.

For an inflationary potential which is flatter than $\phi^{2}$ (such that the scalar self-interaction is attractive), the inflationary condensate may fragment to non-topological solitons. This has notably been examined in the context of fragmentation of Affleck-Dine (AD) condensates in supersymmetric models \cite{kusenko97}, \cite{enqvist98}, \cite{kasuya001}, \cite{kasuya002} , however it is a phenomenon which can also apply to inflation models with appropriate potentials (e.g. hybrid inflation \cite{john02}). This may occur via tachyonic preheating \footnote{For a concise review of tachyonic preheating, see e.g. \cite{kofman01}.} \cite{felder00}, \cite{felder01}, \cite{copeland02}, \cite{john05}, a phenomenon which can occur when there is a tachyonic instability in the potential with respect to spatial perturbations within the inflationary condensate \footnote{For a recent work which considers tachyonic preheating in the context of plateau potentials with a real inflaton, see \cite{tomberg21}.}. If the instability is sufficient, and if there are non-topological soliton solutions present in the theory, then the condensate can fragment into lumps, the precise nature of which depends on the underlying particle physics. However if non-topological soliton solutions do not exist within the theory then the condensate cannot fragment and the perturbations of the inflaton will grow to a maximum within the condensate before being dispersed with expansion, whereupon the inflaton condensate continues its homogeneous evolution as usual. Thus the existence or otherwise of non-topological soliton solutions is essential to the post-inflationary evolution of inflation models with tachyonic preheating.

In a theory with a real scalar inflaton the non-topological solitons are neutral oscillons \footnote{For an in-depth exposition on oscillons and extensive references, see \cite{amin10}. For neutral non-topological solitons whose stability depends on an adiabatic invariant, see \cite{kasuya03}.}. In the case where the scalar field is complex and carries a conserved Noether charge the non-topological solitons will be charged and are called Q-balls\footnote{For an in-depth discussion of  non-topological solitons, see \cite{friedberg76}. For a review, see \cite{lee92}.} \cite{coleman85}. It is these objects that we test for the existence of in the Palatini formalism. The existence of Q-ball solutions suggests that analogous oscillon solutions will also exist for a real inflaton in the Palatini formalism.

The paper is organised as follows. In Section $2$ we introduce the complex Palatini inflation model with an inflaton  mass term and discuss the inflation observables and field strength at the end of inflation. In addition, we derive a new upper bound on the inflaton mass for Palatini inflation to be viable. In Section $3$ we discuss Q-balls for the case of a non-canonically normalised scalar. We derive the Q-ball equation for Palatini inflation in the Einstein frame for the case of a non-canonically normalised complex scalar, which constitutes a new form of the Q-ball equation not previously considered in the literature,  and we discuss the conditions for the stability of the Q-balls. In Section $4$ we determine the range of inflaton mass squared for which Q-balls exist that are compatible with Palatini inflation, which we refer to as the Q-ball window. 
In Section $5$ we present our numerical analysis of Q-ball solutions, where we demonstrate the existence of Q-balls with field strengths comparable to those at the end of Palatini inflation. In Section $6$ we present an analytic approximation to the Q-ball solutions and obtain analytic expressions for the Q-ball radius, energy and charge. In Section $7$ we derive a number of exact relations between the energy and charge of Palatini inflation Q-balls, which generalise existing relations for conventional Q-balls with canonically normalised complex scalars. In Section 8 we consider the possible effects of gravity on our Q-ball solutions. In particular, we determine the conditions under which our flat space Q-ball solutions become black holes. In Section $9$ we discuss the potential significance of our results for (a) the possibility of related oscillon solutions in the case of a real inflaton, (b) the existence of a new class of Q-balls based on complex scalars with non-canonical kinetic terms, (c) inflaton fragmentation and (d) the post-inflation cosmology of Palatini inflation. In Section $10$ we present our conclusions. In Appendix A we give the details of the analysis in Section $7$, and in Appendix B we review the results of the analysis of tachyonic preheating in Palatini inflation of \cite{rubio19}, where we demonstrate that the Q-balls we have obtained are compatible with the size of the non-linear perturbations formed at the end of inflation  derived in \cite{rubio19}.

\section{Palatini Inflation with an Inflaton Mass Term}

We work within the framework of inflation driven by a complex scalar inflaton $\Phi$ charged under a global $U\left(1\right)$ symmetry and non-minimally coupled to gravity. The model we consider differs from the conventional analysis of Palatini inflation, in particular that of Higgs inflation, in that we include a mass term for the inflaton. This requires generalisation of the existing analysis in Palatini inflation. We use the $\left( +, -, -, - \right)$ convention for the metric. $M_{pl}$ denotes the reduced Planck mass.

\noindent The inflaton action in the Jordan frame is given by

\be{e1} 
S_{J} = \int d^{4}x \sqrt{-g} \left[ -\frac{1}{2}M_{pl}^{2}\left( 1 + \frac{2\xi \mid \Phi \mid^{2}}{M_{pl}^{2}}\right)R + \partial_{\mu}\Phi^{ \dagger} \partial^{\mu}\Phi - V\left(\mid \Phi \mid \right) \right] ~.
\ee

\noindent In the Jordan frame the potential is 

\be{ e2 } 
V\left( \mid \Phi \mid \right) = m^{2} \mid \Phi \mid^{2} + \lambda \mid \Phi \mid^{4} 
~,\ee

\noindent where $m$ is the inflaton mass and $\lambda$ is the bare inflaton self-coupling. In order to reformulate the action in the Einstein frame, we perform a conformal transformation on the metric

\be{ e3 } 
g_{\mu \nu} \longrightarrow \tilde{g}_{\mu \nu} = \Omega^{2}g_{\mu \nu} ~,
\ee

\noindent where the conformal factor is

\be{ e4 } 
\Omega^{2} = 1 + \frac{2\xi \mid \Phi \mid^{2}}{M_{pl}^{2}} ~.
\ee

\noindent In the Palatini formulation, the Ricci scalar transforms as 

\be{ e5 } 
R \longrightarrow \tilde{R} = \frac{R}{\Omega^{2}} 
\ee

\noindent and the integration measure transforms as

\be{ e6 } 
\sqrt{-det \left(\frac{\tilde{g}_{\mu\nu}}{\Omega^{2}}\right)} = \sqrt{-\frac{1}{\Omega^{8}}det\left(\tilde{g}_{\mu\nu}\right)} = \frac{1}{\Omega^{4}}\sqrt{-det \tilde{g}_{\mu\nu}} = \frac{\sqrt{-\tilde{g}}}{\Omega^{4}} ~.
\ee
The conformal transformation acts on the metric, so although the derivatives themselves are unaffected, the implicit factor of the metric used in the contraction of the kinetic term is transformed

\be{ e7  } 
\sqrt{-g}\partial_{\mu}\Phi^{\dagger} \partial^{\mu}\Phi = \sqrt{-g}g_{\mu \nu}\partial^{\nu}\Phi^{\dagger} \partial^{\mu}\Phi
\rightarrow 
\sqrt{-\tilde{g}}\frac{\tilde{g}_{\mu \nu}}{\Omega^{2}} \partial^{\nu}\Phi^{\dagger} \partial^{\mu}\Phi =  \sqrt{-\tilde{g}}\frac{1}{\Omega^{2}}\partial_{\mu}\Phi^{\dagger} \partial^{\mu}\Phi ~.
\ee
The Einstein frame action is then

\be{ e8 } 
S_{E} = \int d^{4}x \sqrt{-\tilde{g}} \left[ -\frac{1}{2}M_{pl}^{2}\tilde{R} + \frac{1}{\Omega^{2}}\partial_{\mu}\Phi^{ \dagger} \partial^{\mu}\Phi - V_{E}\left(\mid \Phi \mid \right) \right]
\ee

\noindent where the Einstein frame potential is defined as

\be{e9} 
V_{E}\left(\mid \Phi \mid \right) = \frac{V\left(\mid \Phi \mid \right)}{\Omega^{4}}.
\ee
We can regard the Jordan and Einstein frames as follows: the Jordan frame is the frame in which the model and its symmetries are defined, whereas the Einstein frame is the frame in which the dynamics align with general relativity. We perform all of our calculations in the Einstein frame unless stated otherwise.

\vspace{0.3cm}

\subsection{Slow Roll Parameters and Inflationary Observables}

In this subsection we generalise the calculation of  the inflation observables for non-minimally coupled Palatini inflation to the case with an inflaton mass term. In particular we will show that, although the inflation observables are unaffected by the mass term, there is an upper bound on the inflaton mass for which inflation is possible.

For the purposes of the inflationary calculations we will write the complex field as a radial real field with a complex phase

\be{ e10 } 
\Phi = \frac{\phi}{\sqrt{2}}e^{i\theta}.
\ee

\noindent The conformal factor becomes

\be{ e11 } 
\Omega^{2} = 1 + \frac{\xi \phi^{2}}{M_{pl}^{2}} ~.
\ee

\noindent The scalar potential in the Jordan frame is

\be{ e12 } 
V\left( \phi \right) = \frac{1}{2}m^{2}\phi^{2} + \frac{\lambda}{4}\phi^{4} ~.
\ee

\noindent The Einstein frame potential is then

\be{ e13  } 
V_{E} = \frac{m^{2}\phi^{2}}{2\left( 1 + \frac{\xi \phi^{2}}{M_{pl}^{2}}\right)^{2}} + \frac{\lambda \phi^{4}}{4\left( 1 + \frac{\xi \phi^{2}}{M_{pl}^{2}}\right)^{2}}.
\ee

\noindent Since inflation takes place upon the plateau of the potential, and we expect the field to be very large in this regime, we define what we will refer to as the plateau limit by

\be{ e14  } 
\frac{\xi \phi^{2}}{M_{pl}^{2}} >> 1 ~.
\ee

\noindent This will apply during slow roll inflation. Extracting this factor from the denominators gives

\be{ e15 } 
V_{E} = \frac{\lambda M_{pl}^{4}}{4 \xi^{2}\left( 1 + \frac{M_{pl}^{2}}{\xi \phi^{2}}\right)^{2}}\left[ 1 + \frac{2 m^{2}}{\lambda \phi^{2}}\right].
\ee

\noindent Since $M_{pl}^{2}/\xi \phi^{2} << 1$, we can write to leading order in $M_{pl}^{2}/\xi \phi^{2}$

\be{ e16 } 
V_{E} \thickapprox \frac{\lambda M_{pl}^{4}}{4 \xi^{2}}\left[ 1 + \frac{2 m^{2}}{\lambda \phi^{2}}\right] \left( 1 - 2\frac{M_{pl}^{2}}{\xi \phi^{2}} \right).
\ee
Since $m^{2}/\lambda \phi^{2} <<1$ during inflation, the Einstein frame potential is, to leading order in small terms,
\be{ e17 } 
V_{E} \thickapprox  \frac{\lambda M_{pl}^{4}}{4\xi^{2}}\left[ 1 + \frac{2m^{2}}{\lambda \phi^{2}} - 2\frac{M_{pl}^{2}}{\xi \phi^{2}} \right] = \frac{\lambda M_{pl}^{4}}{4\xi^{2}}\left[ 1 - 2\frac{M_{pl}^{2}}{\xi \phi^{2}}\beta \right]
\ee
where
\be{ e18  } 
\beta = 1 - \frac{\xi m^{2}}{\lambda M_{pl}^{2}}.
\ee
In order to have a minimally coupled inflaton, we perform a field rescaling

\be{ e19 } 
\frac{d\sigma}{d\phi} = \frac{1}{\sqrt{1 + \frac{\xi \phi^{2}}{M_{pl}^{2}}}}.
\ee

\noindent The rescaled inflaton field is then

\be{ e20  } 
\sigma \left(\phi \right) = \frac{M_{pl}}{\sqrt{\xi}}\sinh^{-1} \left(\frac{\sqrt{\xi}}{M_{pl}}\phi\right).
\ee

\noindent and so

\be{e21} 
\phi \left(\sigma \right) = \frac{M_{pl}}{\sqrt{\xi}}\sinh \left(\frac{\sqrt{\xi}}{M_{pl}}\sigma \right) = \frac{M_{pl}}{2\sqrt{\xi}}\left(e^{\frac{\sqrt{\xi}}{M_{pl}}\sigma} - e^{-\frac{\sqrt{\xi}}{M_{pl}}\sigma} \right). 
\ee
It is important to note that for a complex inflaton we cannot in general rescale to a canonical scalar field, in particular when discussing Q-balls. However, for inflation specifically the dynamics are determined by the radial field.

The rescaled field $\sigma$ will be large compared to $M_{Pl}/\sqrt{\xi}$ during inflation, therefore the negative exponential will be significantly smaller than the positive exponential in \eq{e21}. To a good approximation during inflation we therefore have 
\be{ e22  } 
\phi \thickapprox \frac{M_{pl}}{2\sqrt{\xi}} e^{\frac{\sqrt{\xi}}{M_{pl}}\sigma}.
\ee

\noindent Substituting this into the Einstein frame potential gives

\be{ e23 } 
V_{E}\left(\sigma \right) = \frac{\lambda M_{pl}^{4}}{4\xi^{2}}\left[ 1 - 8\beta e^{-2\frac{\sqrt{\xi}}{M_{pl}}\sigma}  \right].
\ee

\noindent The effect of the mass term on the Einstein frame potential is to introduce the factor of $\beta$, whereas $\beta =1$ for conventional massless Palatini inflation. 

We next show that although $\beta$ has no effect on the inflationary observables when expressed in terms of the number of e-folds $N$, it will alter the value of $\phi$.
The slow roll parameters are 
\be{ e24  } 
\epsilon = \frac{M_{pl}^{2}}{2}\left(\frac{V_{E}'}{V_{E}} \right)^{2} \;\;;\;\;
\eta = M_{pl}\frac{V_{E}''}{V_{E}}
\ee
where a prime denotes a derivative with respect to $\phi$. Since $\phi$ is large on the plateau we can approximate $V_{E}$ itself by
\be{ e25 } 
V_{E} \thickapprox \frac{\lambda M_{pl}^{4}}{4\xi^{2}} ~.
\ee
\noindent The derivatives are
\be{ e26 } 
\frac{\partial V_{E}}{\partial \phi} = \frac{\lambda M_{pl}^{4}}{4\xi^{2}}\left(\frac{16\beta \sqrt{\xi}}{M_{pl}}\right) e^{-2\frac{\sqrt{\xi}}{M_{pl}}\sigma} \;\; ; \;\; 
\frac{\partial^{2}V_{E}}{\partial \phi^{2}} =  \frac{\lambda M_{pl}^{4}}{4\xi^{2}}\left(-\frac{32\beta \xi}{M_{pl}^{2}}\right) e^{-2\frac{\sqrt{\xi}}{M_{pl}}\sigma}.
\ee
Calculating the slow roll parameters then gives
\be{ e27 }  
\epsilon = 128 \beta^{2} \xi e^{-4\frac{\sqrt{\xi}}{M_{pl}}\sigma} \;\; ; \;\;
\eta = -32\beta \xi e^{-2\frac{\sqrt{\xi}}{M_{pl}}\sigma}.
\ee
The number of e-folds of inflation is given by
\be{ e28 } 
N = -\frac{1}{M_{pl}^{2}}\int^{\sigma_{end}}_{\sigma} \frac{V_{E}}{V_{E}'} \; d\sigma
 = \frac{1}{32\xi \beta} \left[ \exp \left(\frac{2\sqrt{\xi}}{M_{pl}}\sigma \right) - \exp \left(\frac{2\sqrt{\xi}}{M_{pl}}\sigma_{end} \right)\right]. 
\ee

\noindent Since $\sigma_{end} << \sigma(N)$ for the values of $N$ of interest, $N$ is given by
\be{ e29 } 
N \thickapprox \frac{1}{32\xi \beta}\exp \left(\frac{2\sqrt{\xi}}{M_{pl}}\sigma \right) \Rightarrow \sigma \left(N \right) \thickapprox  \frac{M_{pl}}{2\sqrt{\xi}} \ln \left(32\xi \beta N \right) ~.
\ee
We substitute $\sigma(N)$ to find $\eta$ and $\epsilon$ in terms of the number of e-folds of inflation
\be{ e30 } 
\epsilon = 128 \beta^{2} \xi \exp \left[-4\frac{\sqrt{\xi}}{M_{pl}} \frac{M_{pl}}{2\sqrt{\xi}}\ln \left(32\xi \beta N \right)\right]  =
 \frac{1}{8\xi N^{2}}
\ee
and
\be{ e31 } 
\eta = -32\beta \xi \exp \left[-2\frac{\sqrt{\xi}}{M_{pl}} \frac{M_{pl}}{2\sqrt{\xi}} \ln \left(32\xi \beta N \right)\right]  =  -\frac{1}{N}.
\ee
\noindent These are the standard expressions for the Palatini slow roll parameters and are independent of $\beta$. The scalar spectral index is given by
\be{ e32 } 
n_{s} = 1 + 2\eta - 6\epsilon \thickapprox 1 - \frac{2}{N},
\ee
\noindent the tensor to scalar ratio is 
\be{ e33 } 
r \thickapprox 16\epsilon = \frac{2}{\xi N^{2}}.
\ee
\noindent  and the primordial curvature power spectrum is 
\be{e34} 
P_{\textit{R}} = \frac{V_{E}}{24\pi^{2} \epsilon M_{pl}^{4}} =  \frac{\lambda}{12\xi \pi^{2}} N^{2} ~.
\ee
\noindent The observed amplitude of the power spectrum is $A_{s} = 2.099 \times 10^{-9}$. With $\lambda = 0.1$ and $N = 55$ this gives a value for the non-minimal coupling  $\xi = 1.2163 \times 10^{9}$ and the inflationary observables are
$n_{s} = 0.9636$ and $r = 6.01 \times 10^{-13}$.  The scalar spectral index is within the bounds of the 2018 results from the Planck satellite (assuming $\Lambda$CDM and no running of the spectral index), $n_{s} = 0.9649 \pm 0.0042$ (1-$\sigma$) \cite{planck18}, whilst the tensor to scalar ratio is heavily suppressed. It should be noted that these estimates of the values of the parameters may change in practice since, as we will discuss later, the post-inflationary cosmology of this model may have an effect on the number of e-folds of expansion and consequently on the location of the pivot scale.

$\phi$ can be expressed in terms of the number of e-folds as 
\be{ e35 } 
\phi \left(N \right) = 2\sqrt{2}M_{pl} \sqrt{\beta N}.
\ee
This is dependent upon the inflaton mass term through $\beta$. Slow roll inflation ends when $\mid \eta \mid \approx 1$, giving $N_{end} \approx 1$. The field at the end of inflation is then
\be{ e36  } 
\phi_{end} \approx 2\sqrt{2}M_{pl} \sqrt{\beta}.
\ee
Since $\beta$ will be in the range $0.1-1$ for the models of interest, it is safe to estimate that the inflaton has a value of the order of the Planck scale at the end of inflation. $\beta$ will only have a noticeable effect on the value of the field at the end of inflation for masses very close to the upper limit for which inflation is possible, which we next discuss.

\subsection{The upper bound on the inflaton mass. }

In the following we will be interested in Q-ball solutions which are compatible with Palatini inflation. The existence of Q-ball solutions will be shown to depend strongly upon the inflaton mass.  The range of inflaton mass for which inflation is possible therefore plays an important role in determining the range of inflaton masses for which Q-ball solutions are compatible with Palatini inflation.

In order for the model to inflate successfully, its potential must have a positive gradient with respect to the inflaton. The Einstein frame potential is
\be{ e37 } 
V_{E}\left(\phi \right) = \frac{m^{2} \phi^{2}}{2\Omega^{4}} + \frac{\lambda \phi^{4}}{4\Omega^{4}}.
\ee
\noindent Differentiating the potential gives
\be{ e38 } 
\frac{\partial V_{E}}{\partial \phi} = \frac{1}{\Omega^{6}}\left[m^{2}\phi + \left(\lambda - \frac{\xi m^{2}}{M_{pl}^{2}}\right) \phi^{3} \right].
\ee
In order for inflation to occur, we require that
\be{ e40 } 
\frac{\partial V_{E}}{\partial \phi} > 0    \Rightarrow \lambda - \frac{\xi m^{2}}{M_{pl}^{2}} > 0.
\ee
\noindent From this condition we arrive at the following upper bound on the inflaton mass from the requirement that inflation can occur,
\be{e41} 
m^{2} < \frac{\lambda M_{pl}^{2}}{\xi} ~.
\ee

\noindent 
\noindent

\section{Q-balls}

Q-balls were derived as a natural consequence of a scalar theory with a global $U(1)$ Noether charge \cite{coleman85}. They are a type of non-topological soliton\footnote{For a recent review of non-topological solitons, Q-balls and related objects in $U(1)$ symmetric theories see \cite{nugaev19}.} \cite{friedberg76}, \cite{lee92}, and appear as extended objects composed of scalars with a rotating configuration in field space. These solutions are found by the extremisation of the energy of the field with respect to the corresponding conserved Noether charge, and indeed it is the conserved charge of the field which allows these objects to be stable. Q-balls frequently appear in minimally supersymmetric theories in conjunction with the Affleck-Dine mechanism  \cite{ad} (see as a starting point \cite{kusenko97}), and have been shown to be a possible dark matter candidate (see e.g. \cite{kusenko98}).

In this section we present the derivation of the Q-ball equation in non-minimally coupled Palatini gravity for a complex inflaton charged under a global $U\left(1 \right)$ symmetry\footnote{For a recent study of how to extract the properties of Q-balls when the $U\left( 1 \right)$ symmetry is promoted to a gauge symmetry see \cite{heeck212} and the references therein. For an investigation of the stability of $U(1)$ gauged Q-balls see \cite{smolyakov16}.}.  It is important to emphasise that the non-minimally coupled Palatini action in the Einstein frame cannot be reformulated in terms of a canonical complex scalar field, and therefore requires a different analysis to that traditionally used in the mathematical treatment of Q-balls. This results in a new class of Q-balls for which, in the Jordan frame, the scalar self-interaction arises due to the non-minimal coupling to gravity, and where in the Einstein frame the scalar field has a non-canonical kinetic term\footnote{A numerical study of the formation of non-minimally coupled metric Q-balls in a SUSY flat direction Affleck-Dine framework, and the consequences for the gravitational wave spectrum was released while this work was in progress \cite{wang21}.}.

\subsection{Derivation of the Q-ball equation for non-canonically normalised complex scalars }

One of the primary objectives in this work is to establish the existence of Q-balls in Palatini gravity in conjunction with successful inflation. We begin here by deriving the Q-ball equation for the non-canonically normalised field in the Einstein frame. We work in flat space with the action

\be{ e42  } 
S = \int d^{4}x \; \frac{1}{\Omega^{2}}\partial_{\mu}\Phi^{\dagger} \partial^{\mu}\Phi - \frac{1}{\Omega^{4}}V\left(\mid \Phi \mid \right) ~,
\ee

\noindent  Gravitational effects can affect the stability of Q-balls or alter the size predictions (see e.g. \cite{multamaki022}, \cite{tamaki11}). We will consider the possible gravitational effects on our Q-ball solutions once we have determined their flat space properties.

We use the method of Lagrange multipliers in order to minimise the energy of the condensate for a fixed charge. The Q-ball energy functional is \footnote{Note that this is a mathematical object and not a physical energy. However, it reduces to the physical energy $E$ when $\Phi$ is a Q-ball solution, since the $\omega$ term then vanishes.} 

\be{ e43  } 
E_{Q} = E + \omega \left( Q - \int d^{3}x  \rho_{Q} \right) ~.
\ee

\noindent The global energy and charge are

\be{ e44 } 
Q = \int d^{3}x \; j^{0} = \int d^{3}x \, \,  \rho_{Q} ~
\ee
and
\be{ e45 } 
E = \int d^{3} \; x T^{00} = \int d^{3}x\, \,   \rho_{E} ~,
\ee

\noindent where $j^{0}$ is the temporal component of the conserved $U(1)$ Noether current $j^{\mu}$, and $T^{00}$ is the temporal component of the energy-momentum tensor $T^{\mu \nu}$. The energy-momentum tensor is given by

\be{ e46 } 
T^{\mu \nu} = \frac{\partial \mathcal{L}}{\partial \left(\partial_{\mu}\phi_{a}\right)}\eta^{\nu \rho}\partial_{\rho}\phi_{a} - \delta^{\mu}_{\rho} \eta^{\nu \rho}\mathcal{L} ~,
\ee

\noindent where $\eta^{\nu \rho}$ is the Minkowski metric. The temporal component is given by

\be{e47} 
\rho_{E} = T^{00} = \frac{1}{\Omega^{2}}\partial_{t}\Phi^{\dagger} \partial_{t}\Phi + \frac{1}{\Omega^{2}}\partial_{i}\Phi^{\dagger} \partial_{i}\Phi + \frac{V(\mid \Phi \mid)}{\Omega^{4}}.
\ee

\noindent We derive the charge density from the conserved Noether current $j^{\mu}$ of the $U\left(1 \right)$ symmetry of the model, where

\be{ e48 } 
j^{\mu} = \frac{\partial \mathcal{L}}{\partial \left(\partial_{\mu}\phi_{a}\right)}\delta \phi_{a} ~,
\ee

\noindent and $\partial_{\mu} j^{\mu} = 0$. In this model the temporal component of the conserved Noether current is

\be{ e49 } 
j^{0} = \frac{\partial \mathcal{L}}{\partial \left(\partial_{t}\Phi \right)} i\Phi - \frac{\partial \mathcal{L}}{\partial \left(\partial_{t} \Phi^{\dagger}\right)} i\Phi^{\dagger},
\ee

\noindent which gives the charge density 

\be{e50} 
\rho_{Q} = \frac{i}{\Omega^{2}}\left( \Phi \partial_{t}\Phi^{\dagger} - \Phi^{\dagger}\partial_{t}\Phi \right).
\ee
Substituting \eq{e47} and \eq{e50} into the Q-ball energy functional gives
\be{ e51  } 
E_{Q} = \int d^{3}x \left[ \frac{1}{\Omega^{2}}\partial_{t}\Phi^{\dagger} \partial_{t}\Phi + \frac{1}{\Omega^{2}}\partial_{i}\Phi^{\dagger} \partial^{i}\Phi + \frac{V(\mid \Phi \mid)}{\Omega^{4}} - \frac{\omega i}{\Omega^{2}}\left( \Phi \partial_{t}\Phi^{\dagger} - \Phi^{\dagger}\partial_{t}\Phi \right) \right] + \omega Q  ~,
\ee

\noindent which can be written as

\be{ e52 } 
E_{Q} = \int d^{3}x \left[ \frac{1}{\Omega^{2}} \mid \partial_{t}\Phi - i \omega \Phi \mid^{2} - \frac{1}{\Omega^{2}} \omega^{2} \mid \Phi \mid^{2} + \frac{1}{\Omega^{2}}\partial_{i}\Phi ^{\dagger} \partial^{i} \Phi + \frac{1}{\Omega^{4}}V\left(\mid \Phi \mid \right) \right] + \omega Q ~.
\ee

\noindent In order to extremise $E_{Q}$ we therefore require that

\be{ e53 } 
\Phi \left(x,t \right) = \Phi \left( x \right) e^{i \omega t} ~.
\ee

\noindent This gives

\be{ e54 } 
E_{Q} = \int d^{3}x \left[ \frac{1}{\Omega^{2}} \mid \overrightarrow{\nabla} \Phi \mid^{2} + \frac{1}{\Omega^{4}} V\left(\mid \Phi \mid \right) - \frac{1}{\Omega^{2}}\omega^{2} \mid \Phi \mid^{2} \right].
\ee

\noindent We define

\be{e55} 
V_{\omega} \left( \mid \Phi \mid \right) = \frac{1}{\Omega^{4}}V \left( \mid \Phi \mid \right) - \frac{1}{\Omega^{2}}\omega^{2}\mid \Phi \mid^{2}
\ee

\noindent as the Q-ball potential. Let us now assume that the Q-balls are spherically symmetric to minimise the energy of the field (without loss of generality we can assume that $\Phi(x)$ is real) 

\be{ e56 } 
\Phi \left(x\right) = \frac{\phi\left(r\right) }{\sqrt{2}} ~.
\ee

\noindent Recasting the integral into spherical polar coordinates, we have

\be{e57} 
E_{Q} = \int dr \;  4\pi r^{2} \left[ \frac{1}{2\Omega^{2}}\left(\frac{\partial \phi}{\partial r}\right)^{2} + V_{\omega} \left(\phi \right) \right] + \omega Q.
\ee

\noindent We can define an effective Lagrangian ${\cal L}_{Q}$,  where  

\be{e58} 
E_{Q} = \int dr \; \mathcal{L}_{Q} + \omega Q ~,
\ee

\noindent  and  

\be{ e59 } 
\mathcal{L}_{Q}= 4\pi r^{2} \left[ \frac{1}{2\Omega^{2}}\left(\frac{\partial \phi}{\partial r}\right)^{2} + V_{\omega} \left(\phi \right) \right] ~.
\ee

\noindent We then apply the Euler-Lagrange equations to extremise $E_{Q}$, 

\be{ e60 } 
\frac{\partial \mathcal{L}_{Q}}{\partial \phi} - \frac{d}{dr}\left(\frac{\partial \mathcal{L}_{Q}}{\partial \left(\partial_{r}\phi\right)}\right) = 0.
\ee

\noindent where  

\be{ e61 } 
\frac{d}{dr}\left(\frac{\partial \mathcal{L}_{Q}}{\partial \left(\partial_{r}\phi\right)}\right) = \frac{8\pi r}{\Omega^{2}}\frac{\partial \phi}{\partial r} + \frac{4 \pi r^{2}}{\Omega^{2}}\frac{\partial^{2}\phi}{\partial r^{2}} - \frac{8 \pi r^{2}}{M_{pl}^{2}}\frac{\xi \phi}{\Omega^{4}}\left(\frac{\partial \phi}{\partial r}\right)^{2}
\ee

\noindent and

\be{ e62  } 
\frac{\partial \mathcal{L}_{Q}}{\partial \phi} = 4\pi r^{2} \frac{\partial V_{\omega}}{\partial \phi} - \frac{4 \pi r^{2}}{M_{pl}^{2}}\frac{\xi \phi}{\Omega^{4}}\left(\frac{\partial \phi}{\partial r}\right)^{2}.
\ee

\noindent Substituting these and and multiplying by $\Omega^{2}/4 \pi r^2$ gives the Q-ball equation for the non-minimally coupled complex scalar

\be{e63} 
\frac{\partial^{2}\phi}{\partial r^{2}} + \frac{2}{r}\frac{\partial \phi}{\partial r} - K\left(\phi \right) \left(\frac{\partial \phi}{\partial r}\right)^{2} = \Omega^{2}\frac{\partial V_{\omega}}{\partial \phi} ~,
\ee

\noindent where 

\be{ e64  } 
K(\phi) = \frac{\xi \phi}{M_{pl}^{2} \Omega^{2}}
\ee

\noindent and 

\be{e65}
V_{\omega} = \frac{V\left(\phi \right)}{\Omega^{4}}  - \frac{\omega^{2}\phi^{2}}{2\Omega^{2}}.
\ee

\noindent This is a new form of the Q-ball equation as  compared to that seen in a scalar theory with canonical kinetic terms, and therefore defines a new class of Q-ball. Setting $\xi = 0$ and $\Omega^{2} = 1$ recovers the result of the conventional Q-ball equation for the minimally coupled case

\be{e66} 
\frac{\partial^{2} \phi}{\partial r^{2}} + \frac{2}{r}\frac{\partial \phi}{\partial r} = \frac{\partial V_{\omega}}{\partial \phi}
\ee

\noindent where in this instance

\be{ e67  } 
V_{\omega} \left( \phi \right) = V(\phi) - \frac{1}{2}\omega^{2}\phi^{2}.
\ee

\subsection{The non-canonical Q-ball equation in terms of a quasi-canonically normalised scalar} 

Unlike the case of a real scalar, it is not possible to transform the complex field $\Phi$ in the Einstein frame to a canonically normalised complex scalar. However, once we have determined the time dependence of the Q-ball solution, it is possible to transform the radial field $\phi(r)$ to a field that resembles a canonically normalised scalar.
As we will now show, the Palatini Q-ball equation for a non-canonically normalised scalar can be written in terms of a non-physical rescaled field, $\sigma$, such that the Q-ball equation takes almost the same form as the conventional Q-ball equation for minimally coupled scalars, and therefore the solutions can be understood as being similar.

The non-canonical Q-ball equation is 

\be{ e68 } 
\frac{\partial^{2} \phi}{\partial r^{2}} + \frac{2}{r}\frac{\partial \phi}{\partial r} - \frac{\xi \phi}{M_{pl}^{2} \Omega^{2}} \left(\frac{\partial \phi}{\partial r}\right)^{2} = \Omega^{2}\frac{\partial V_{\omega}}{\partial \phi}.
\ee

\noindent We transform this according to the rescaling

\be{e69} 
\frac{d\phi}{d r} = \Omega \frac{d\sigma}{d r} \;\;;\;\; \Omega^{2} = 1 + \frac{\xi \phi^{2}(r)}{M_{Pl}^{2}}  ~.
\ee

\noindent Therefore

\be{ e70  } 
\frac{d^{2}\phi}{d r^{2}} = \Omega \frac{d^{2}\sigma}{d r^{2}} + \frac{d\Omega}{d r }\frac{d\sigma}{d r}
\ee

\noindent and 

\be{ e71  } 
\frac{d V_{\omega}}{d\phi} = \frac{1}{\Omega}\frac{d V_{\omega}}{d \sigma}.
\ee

\noindent The rescaled Q-ball equation is then

\be{e72} 
\frac{d^{2}\sigma}{d r^{2}} + \frac{2}{r}\frac{d\sigma}{d r} = \frac{d V_{\omega}}{d\sigma} ~,
\ee

\noindent where $V_{\omega}$ is given by \eq{e65}.

\noindent It is important to emphasize that while this equation is similar to the conventional Q-ball equation for a canonically normalised complex scalar, it is not the same. In the canonically normalised case we would have
\be{ e74  } 
V_{\omega} = \frac{V\left(\phi\left(\sigma \right) \right)}{\Omega^{4}} - \frac{\omega^{2}\sigma^{2}}{2}.
\ee
Therefore $\sigma^{2} \rightarrow \phi^{2}(\sigma)/\Omega^{2}(\sigma)$ in the non-canonically normalised case. This will result in significantly different Q-ball solutions, but it also suggests that we can expect a broad similarity to the conventional case.

\subsection{The mechanical analogy and the condition on $\omega$ for existence of Q-ball solutions}

The existence of Q-ball solutions in a given theory is heavily dependent upon the form of the potential, and indeed the nature of the complex scalar. There is a useful mechanical analogy \cite{coleman85} which can be used to understand the precise nature of Q-ball solutions and the existence conditions behind them. This analogy was originally derived for Q-balls in the minimally coupled case, therefore the underlying dynamics in the present model will be different.

We first consider the conventional case of a canonically normalised scalar,  for which $\sigma = \phi$. If we interpret the field $\sigma$ as a position coordinate and $r$ as a time coordinate, then \eq{e72} looks like it is describing the equation of motion of a particle moving in the potential $-V_{\omega}$ with a damping term. A schematic of the potential in this model with the symmetric minima shown is illustrated in Figure $1$.

\begin{figure}
\begin{center}
\includegraphics[clip = true, width=0.75\textwidth, angle = 360]{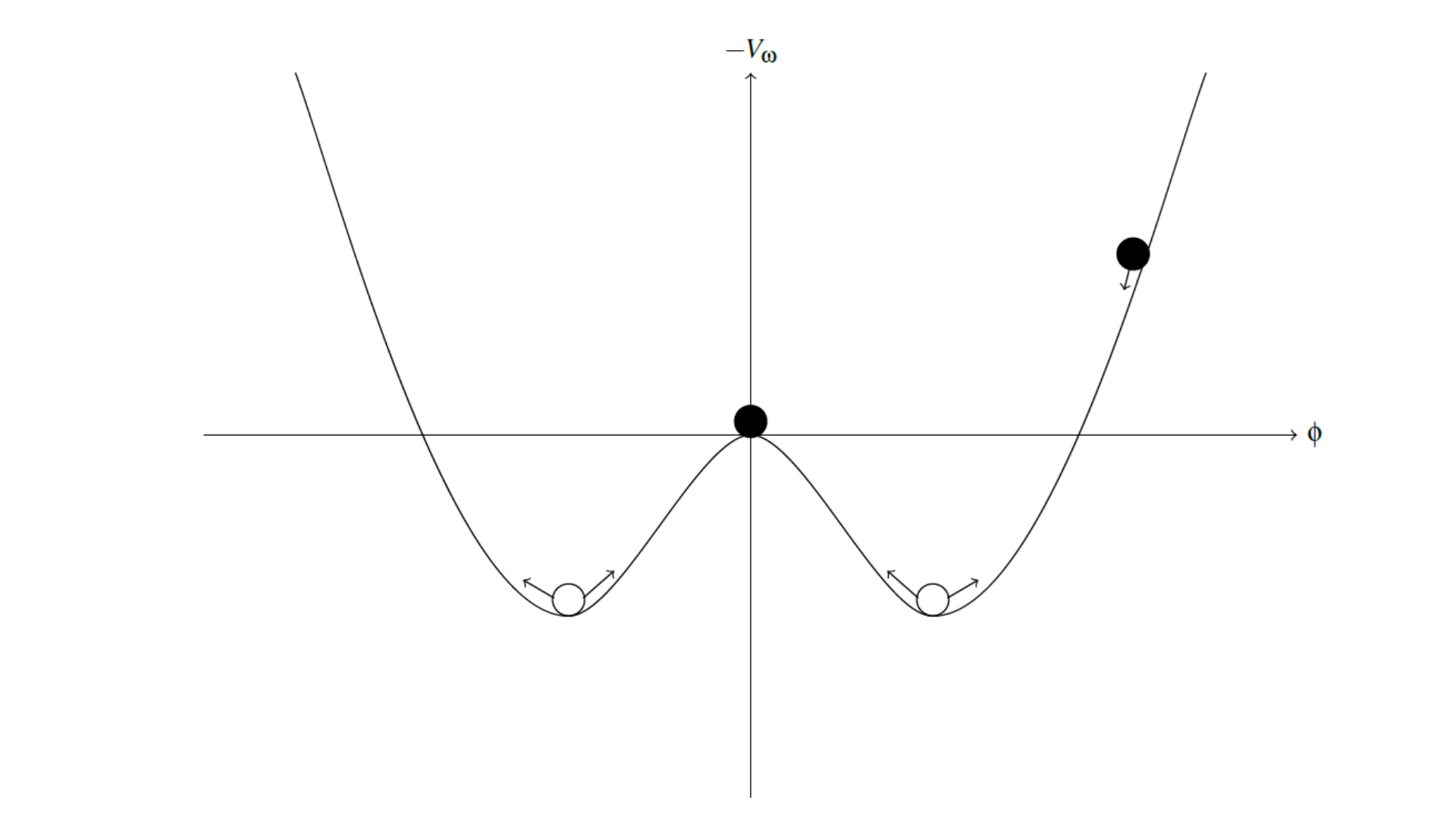}
\caption{Schematic of the negative potential of the mechanical analogy showing the symmetric minima.} 
\label{fig1}
\end{center}
\end{figure}

\begin{figure}[H]
\begin{center}
\includegraphics[clip = true, width=0.75\textwidth, angle = 360]{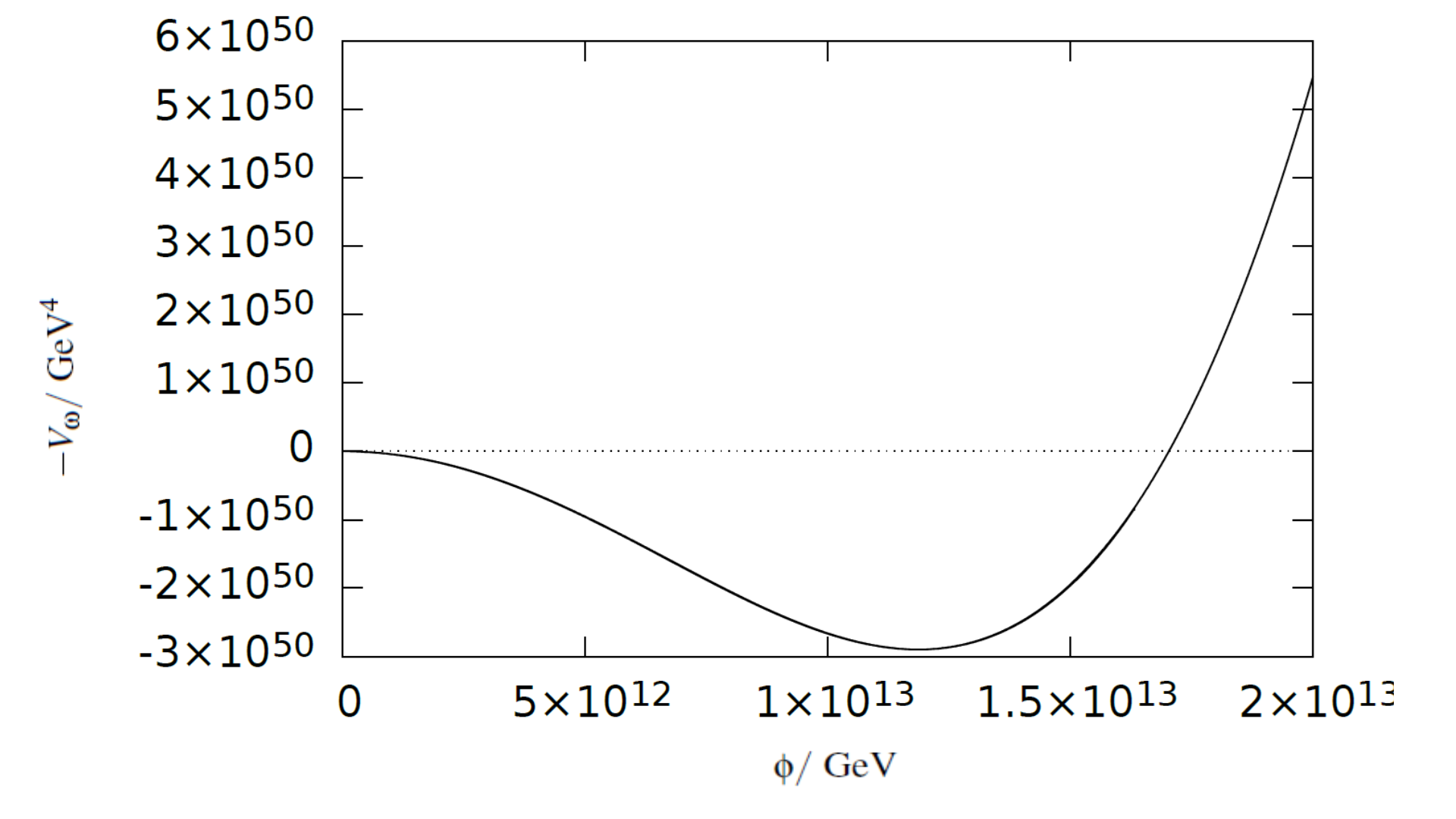}
\caption{$-V_{\omega}(\phi)$ for a large $\omega$ Q-ball in our non-minimally coupled Palatini model.} 
\label{fig2}
\end{center}
\end{figure}

\begin{figure}[H]
\begin{center}
\includegraphics[clip = true, width=0.75\textwidth, angle = 360]{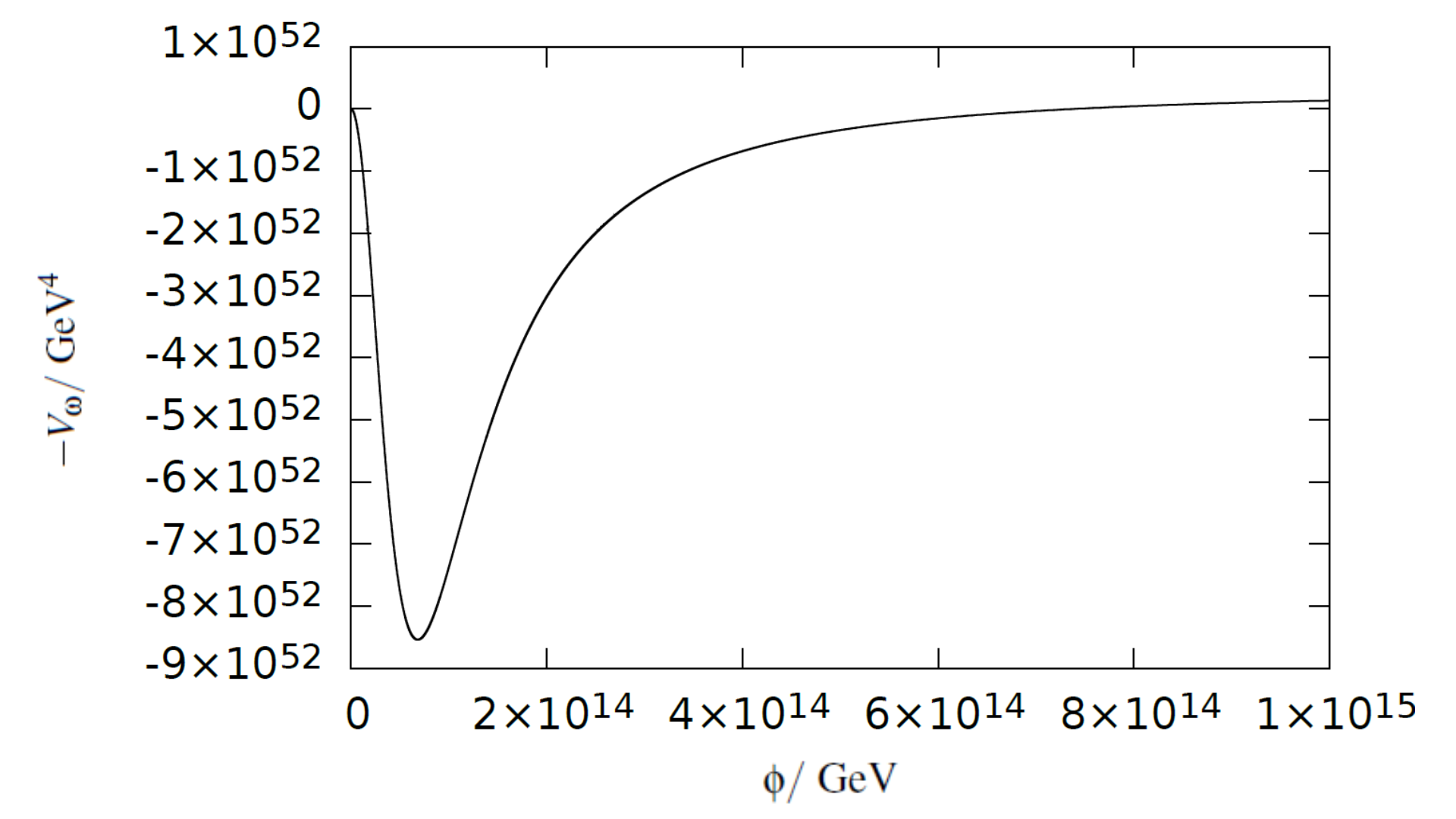}
\caption{$-V_{\omega}(\phi)$ for a small $\omega$ Q-ball in our non-minimally coupled Palatini model.} 
\label{fig3}
\end{center}
\end{figure}

The "particle" begins its motion from the right (black circle), and travels down its potential from some position $\phi_{0}$. The local maximum at the origin of the potential $\phi = 0$ corresponds to the field in the vacuum at $r \rightarrow \infty$; this is where we need the field to come to rest in order for the theory to admit Q-ball solutions. There are a number of requirements for this to be realised. Firstly there is the matter of the starting "position" of the particle ($\phi_{0}$) necessary in order for the particle to come to rest exactly at the origin. If the particle begins its motion too close to the origin of its potential i.e. $\phi_{0}$ is too small, or the gradient of the potential is too shallow, then the particle will not have enough momentum to reach the top of the hill at the origin and it will end up oscillating in the local minimum to the right (clear circle on the right hand side), corresponding to undershoot. Place the starting point too far away, or have a potential that is too steep on the approach to the positive $\phi$ minimum, and the field will be rolling too fast and will roll through the origin over to the other side (clear circle on the left hand side), corresponding to overshoot.

This is still a good analogy for the non-canonically normalised case, since the potential takes the same form. (Plots of $-V_{\omega}$ for positive $\phi$ and different $\omega$ are shown in Figures $2$ and $3$.)
In terms of $\phi$ the analogy is different to the conventional case and the dynamics are less straightforward. In the non-minimally coupled Palatini case, the Q-ball equation in terms of $\phi$, \eq{e63}, has, in addition to a factor of $\Omega^2$ multiplying the derivative of the potential, an additional gradient squared term with a negative sign. This can be interpreted as an external energy input to the field - an 'anti-friction' if you will - which changes the motion of the particle from the straightforward Newtonian analogy applied in the conventional case. In terms of $\sigma$, the Q-ball equation, \eq{e72}, has the same form as in the conventional case but with a modified potential $V_{\omega}$, \eq{e65}. We will therefore consider the mechanical analogy using this equation, in which $\phi$ is a function of $\sigma$.

We next derive the bounds on $\omega$ to ensure that the potential is compatible with the existence of Q-ball solutions; later we will demonstrate numerically some examples of Q-balls with compatible $\omega$ and $\phi_{0}$ in this framework.

In order to avoid undershoot, the starting point, $-V_{\omega}\left(\phi_{0}\right)$, needs to sit above, or at least at zero, in order for the particle to gather enough momentum to reach the local maximum of the origin. We can write this as

\be{ e75 } 
max\left(-V_{\omega}\right) \geq 0.
\ee

\noindent This gives a lower bound on $\omega$ in the standard theory. In order to avoid overshoot we require that the origin is a local maximum so that the particle doesn't roll into the negative field region, such that the particle can come to rest at $\phi = 0$ (noting that $\phi$ is equivalent to $\sigma$ as $\phi \rightarrow 0$)  

\be{ e76  } 
\left. \frac{d^{2}\left(-V_{\omega} \right)}{d\phi^{2}}\right|_{\phi = 0} < 0 \;\; \Rightarrow 
\omega^{2} < \left. \frac{d^{2}V}{d\phi^{2}} \right|_{\phi = 0}.  
\ee

\noindent In the limit $\phi \rightarrow 0$ we have $\Omega \rightarrow 1$ and

\be{ e77  }  
\frac{d^{2}\left(-V_{\omega} \right)}{d\phi^{2}} \rightarrow \omega^{2} - m^{2} - 3\lambda \phi^{2}  ~.
\ee

\noindent  
\noindent Therefore

\be{ e78 } 
\left. \frac{d^{2}\left(-V_{\omega} \right)}{d\phi^{2}}\right|_{\phi = 0} = 0 \Rightarrow \omega^{2} - m^{2} < 0 ~ .
\ee

\noindent  Thus the existence condition for Q-balls is

\be{e79} 
\omega < m  ~.
\ee

\noindent This the same result as that for the existence of conventional canonically normalised Q-balls.

\subsection{Q-ball Stability}

There are a number of different types of Q-ball stability. Here we will concern ourselves with so-called \textit{absolute stability} although there are other criteria for Q-ball stability, in particular \textit{classical stability}.
A Q-ball is said to be absolutely stable if \cite{coleman85}

\be{ e80  } 
E < mQ,
\ee

\noindent where  $E$ is the energy of the Q-ball. $mQ$ can be interpreted as the energy of $Q$ free quanta of scalar particles of mass $m$ in the vacuum. This simply states that the energy of a Q-ball must be less than the sum of its parts in order to be absolutely stable. The energy difference is the binding energy of the Q-ball.
In contrast, classical stability implies a solution that is stable against small perturbations. The classical stability of a Q-ball can be gauged by the sign of the derivative of the charge with respect to $\omega$, that is if $\partial Q/\partial \omega <0$ then a Q-ball is classically stable \cite{friedberg76} \cite{lee92}.

\section{ The allowed range of inflaton mass squared from the existence of Q-balls and inflation }

In this section we will show that there is a range of inflaton mass squared for which Q-balls exist that are compatible with inflation, which we refer to as the Q-ball window.

\subsection{Constraint from the existence of Q-balls}

The Q-ball potential is given by

\be{e81} 
V_{\omega} \left(\phi \right) =  \frac{1}{\Omega^{4}}\left(\frac{1}{2} m^{2} \phi^{2} + \frac{\lambda}{4} \phi^{4} \right)  - \frac{\omega^{2} \phi^{2}}{2\Omega^{2}}.
\ee

\noindent Working to leading order in $M_{pl}^{2}/\xi \phi^{2}$, \eq{e81}  can be expanded as

\be{ e82  } 
V_{\omega} \left(\phi \right) = \frac{M_{pl}^{4}}{2\xi^{2}\phi^{2}}\left( m^{2} + \omega^{2} - \frac{2m^{2}M_{pl}^{2}}{\xi \phi^{2}} \right) - \frac{\lambda M_{pl}^{6}}{2 \xi^{3}\phi^{2}} + \frac{\lambda M_{pl}^{4}}{4\xi^{2}} - \frac{\omega^{2}M_{pl}^{2}}{2\xi} ~.
\ee

\noindent Therefore 

\be{e83} 
V_{\omega}\left(\phi \right) = \frac{M_{pl}^{4}}{2\xi^{2}\phi^{2}}\left[ m^{2} + \omega^{2} - \frac{\lambda M_{pl}^{2}}{\xi} \right] + \hspace{1mm}higher\hspace{1.0mm} order\hspace{1.0mm} and \hspace{1.0mm} constant ~\hspace{0.8mm} terms.
\ee

\noindent  In order to have a Q-ball solution to \eq{e63}, the solution $\phi \left(r \right)$ must decrease as $r$ increases from zero. This requires that that the $\phi$ dependent term on the RHS of \eq{e83} is positive. Therefore the existence condition for a Q-ball is 

\be{e84} 
m^{2} + \omega^{2} > \frac{\lambda M_{pl}^{2}}{\xi}. 
\ee

\noindent In the next section we show analytically that if $\omega < m$, as required to have Q-ball solutions, then the condition \eq{e84} must be satisfied for zeros of the RHS of the Q-ball equation with $\phi \neq 0$ to exist, which is a necessary condition to have Q-ball solutions. 

\noindent Defining

\be{e85} 
\omega_{c}^{2} = \frac{\lambda M_{pl}^{2}}{\xi}
\ee

\noindent we can rewrite \eq{e84} as

\be{e86} 
m^{2} + \omega^{2} > \omega_{c}^{2}. 
\ee

\noindent Combining $\eq{e86}$ with the condition \eq{e41} for inflation to be possible, which may be written as $m^{2} < \omega_{c}^{2}$, we find the necessary condition for Q-balls solutions to exist which are compatible with inflation

\be{e87} 
m^{2} < \omega_{c}^{2} < m^{2} + \omega^{2}.
\ee

\noindent In order for Q-balls to exist,  we also require that $\omega < m$, \eq{e79}, is satisfied. Therefore

\be{ e88 } 
\omega^{2} + m^{2} < 2m^{2}.
\ee

\noindent With this, we can state the range of $\omega_{c}^{2}$  for which stable Q-balls consistent with inflation exist, \eq{e87}, in terms of the inflaton mass squared

\be{ e89 } 
m^{2} < \omega_{c}^{2} < 2m^{2}. 
\ee

\noindent This can equivalently be written as a range of inflaton mass squared

\be{e90} 
\frac{\omega_{c}^{2}}{2} < m^{2} < \omega_{c}^{2}.
\ee

\noindent This gives the window in inflaton mass squared (the Q-ball window) over which we can search for Q-ball solutions which are compatible with inflation. It is interesting to note that the Q-ball window favours inflaton masses close to the upper bound from inflation. It is also possible for Q-balls to exist for masses larger than the upper bound we have here, but the model cannot inflate in this limit.

\section{Numerical Q-ball Solutions}

In this section we numerically solve the Q-ball equation in terms of $\phi$, \eq{e63}, for the following boundary conditions at $r=0$

\be{ e91 } 
\phi \left(r =0\right) = \phi_{0} \;\;\;\; ; \;\;\;
\frac{\partial \phi}{\partial r} \left(r=0\right) = 0.
\ee

\subsection{Zeroes of the Q-ball equation}

In order to reduce the size of the parameter space for $\phi_{0}$ for which we need to search for a Q-ball, it is useful to consider the fixed points corresponding to the zeroes of the RHS of the Q-ball equation \eq{e63}. The fixed points of the equation correlate with the possible behaviours of the mechanical analogy, with the fixed points corresponding to the maxima and minima of the Q-ball effective potential.
Our Q-ball equation \eq{e63} has three zeroes, which correspond to two stable attractor fixed points and one unstable repulsor fixed point. The attractors are symmetric, with one positive $(\phi_{+})$ and one negative $(\phi_{-})$.

\vspace{-0.3cm}

\begin{figure}[H]
\begin{center}
\includegraphics[clip = true, width=0.75\textwidth, angle = 360]{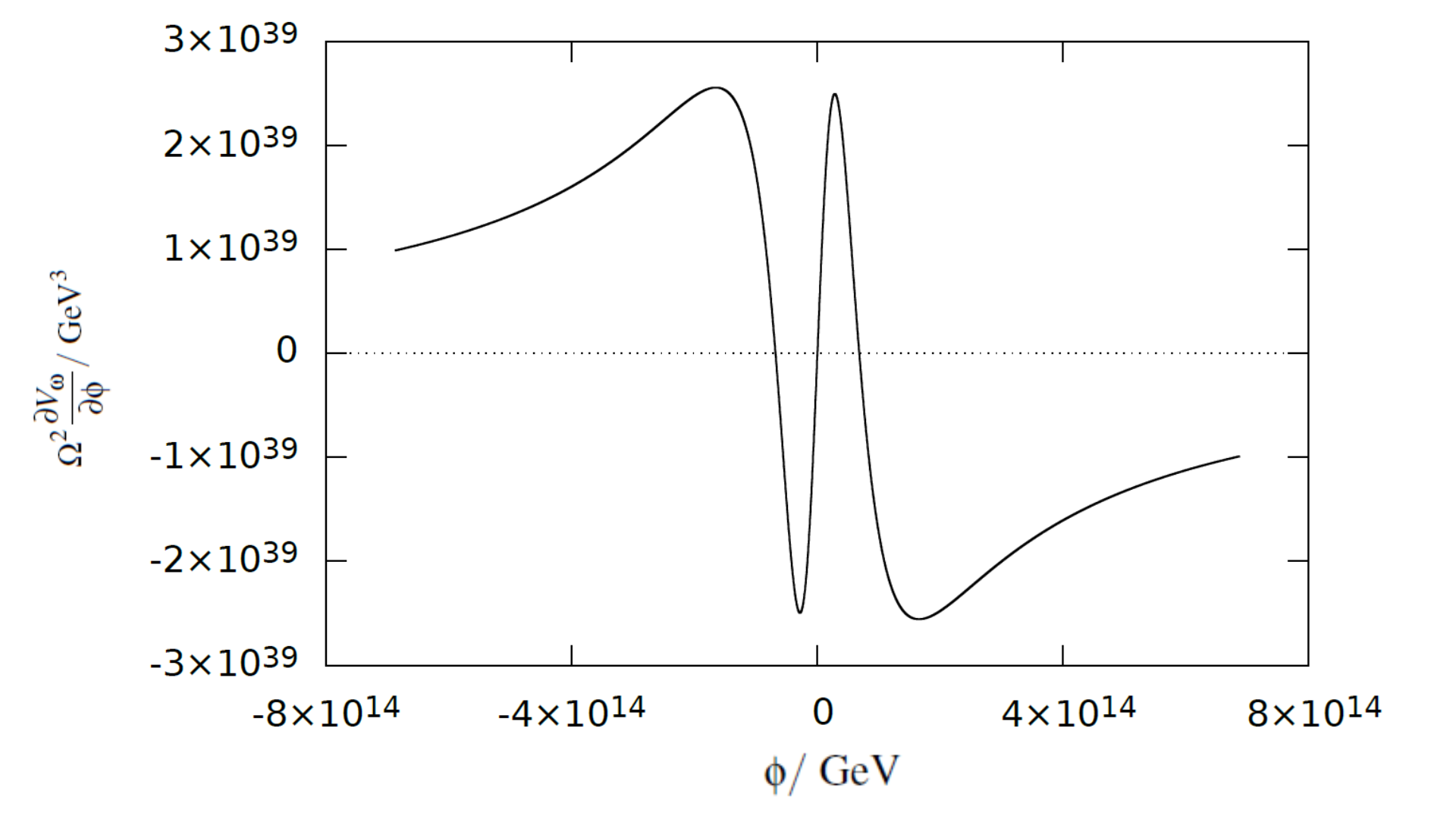}
\caption{Plot illustrating the zeroes of the Q-ball equation.} 
\label{fig4}
\end{center}
\end{figure} 

Figure $4$ shows an example of the right hand side of the Q-ball equation as a function of $\phi$. The points where the function crosses zero are the fixed points of the equation. The attractor fixed points provide a useful reference for determining the value of $\phi$ at $r = 0$, $\phi_{0}$, for which a Q-ball solution exists. The value of $\phi_{0}$ which results in the field getting caught by - and oscillating around -  the positive fixed point $\phi_{+}$ as $r$ increases acts as a lower bound on the value of $\phi_{0}$ which could produce a Q-ball. Similarly, if one pushes $\phi_{0}$ higher, the field can drop below zero as $r$ increases and begin to oscillate around the negative fixed point $\phi_{-}$. This $\phi_{0}$ can then act as an upper bound on the range of $\phi_{0}$ which could produce a Q-ball, with the true $\phi_{0}$ being between the two bounds.
This can also be understood in terms of the mechanical analogy of Q-balls we outlined in Section $3$. The fixed points of the Q-ball equation can be interpreted as the stationary points of the potential $-V_{\omega}$ in the analogy, whereupon a particle rolling down the potential $-V_{\omega}$ can come to rest under its own motion. The first local minimum on the right hand side of the origin in Figure $1$ corresponds to the positive attractor fixed point (the first zero from the right in Figure $4$). The unstable fixed point is at the origin in Figure $4$, and corresponds to the local maximum at $\phi =0$ in Figure $1$. The local minimum at negative $\phi$ in Figure $1$ corresponds to the symmetric negative fixed point in Figure $4$ (furthest zero from the right).

This representation of the zeroes of the Q-ball equation combined with the mechanical analogy illustrates how precise the conditions needed to obtain a Q-ball solution are. The solution corresponding to Q-balls appears as $\phi$ asymptotes to zero as $r \rightarrow \infty$. In the analogy this corresponds to the field coming to rest exactly at the top of the potential at the origin, and having the correct energy and momentum to do so. It also illustrates that the existence of the fixed points themselves determine the existence of Q-ball solutions.

We next analytically derive the condition for the existence of zeros of the Q-ball equation. This requires that 
$\partial V_{\omega}/\partial \phi = 0$ at $\phi \neq 0$. Applying this to the potential \eq{e81} gives the equation for the zeros, 
\be{z1} \phi \left(m^{2} - \omega^{2} - \frac{\xi}{M_{Pl}^{2}} \left(m^{2} + \omega^{2} -\frac{\lambda M_{Pl}^{2}}{\xi}\right) \phi^{2} \right) = 0 ~.\ee 

\noindent The zeros with $\phi \neq 0$ are found at  

\be{z2} \phi = \pm \frac{M_{Pl}}{\sqrt{\xi}} \frac{\sqrt{m^{2} - \omega^{2}}}{\sqrt{m^{2} + \omega^{2} - \omega_{c}^{2} }  }     ~.\ee 

\noindent To have a stable Q-ball solution we require that $\omega^{2} < m^{2}$. Therefore the range of values of $m^{2}$ for which relevant zeros of the Q-ball equation exist is 

\be{z3}    m^{2} + \omega^{2} > \omega_{c}^{2}  ~.  \ee 

\noindent This is the same as the condition \eq{e86} obtained directly from the Q-ball equation. We have confirmed numerically that  zeros with $\phi \neq 0$ indeed exist over the range of $\omega$ satisfying $\omega^2 < m^2$ and $m^{2} + \omega^{2} > \omega_{c}^{2}.$

\vspace{-0.5cm}

\subsection{Results and Discussion}

We next present examples of numerical solutions to \eq{e63} for Palatini Q-balls. Later we will compare these solutions to an approximate analytical solution in order to understand their properties. 

For a fixed $m$ and for a range of values of $\omega$ (both expressed in terms of $\omega_{c}$), we scan the parameter space of $\phi_{0}$ to find a Q-ball solution to six decimal places in $\phi_{0}$. The examples of $\omega$ are chosen to be compatible with the constraints on $m$ and $\omega$ from inflation and the existence of Q-balls (the Q-ball window, \eq{e90}) and to produce a wide range of Q-ball sizes. An ideal Q-ball profile will depict the field decreasing with $r$ from its initial value $\phi_{0}$ at $r=0$ until it asymptotes to zero along the $r$ axis. Scanning for Q-ball solutions of this form is an established technique, and we utilise this method to find ten Q-ball solutions for the case $\lambda = 0.1$ and $m=0.9\omega_{c}$ in the energy range $\mathcal{O}\left(10^{13}\right) - \mathcal{O}\left(10^{17}\right) \GeV$, which represents a typical value of $m^2 (= 0.81 \omega_{c}^{2})$ within the range allowed by the Q-ball window, $ m^{2} = (0.5-1)\omega_{c}^{2}$.

Tables 1 and 2 show the values of $\omega$ and the corresponding $\phi_{0}$ needed to produce a Q-ball, as well as two measures of the Q-ball radius which will be defined shortly. The largest value of $\omega$ ($0.89\omega_{c}$) is intentionally close to the upper bound  from the existence of Q-balls, $m^{2} < \omega_{c}^{2}$. The smaller values of $\omega$ are taken to a very large number of decimal places. This is because when scanning for Q-ball solutions we found that the closer $\omega$ was to the vicinity of $0.7\omega_{c}$ and the larger $\phi_{0}$ became, more sensitive to the initial conditions the numerical solution became. The example quoted in the tables, $0.707155\omega_{c}$, is the smallest value of $\omega$ we obtained, corresponding to the largest $\phi_{0}$ Q-ball; the larger number of decimal points in this result indicates the precise nature of this apparent lower bound on $\omega$ and the sensitivity of the existence of corresponding Q-ball solutions to the model parameters. It is interesting to note that this number is very close to being exactly $\omega_{c}/\sqrt{2}$. It is likely that the emergence of this lower bound on $\omega$ has some theoretical significance, however we have not determined as of yet that there is a way to derive this lower bound analytically although it is a result worth investigating further\footnote{It is worth remarking that this result is not, as far as the authors are aware, reproducible using the standard lower bound on $\omega$ from the condition for avoiding undershoot in the Coleman framework of Q-balls, as in this case we are considering a plateau potential.}. 

Two different tables of values of the parameters of the Q-balls are presented. This is because there is no fixed definition of the radius of a Q-ball, since where exactly its edge lies is up to interpretation. We have taken two different definitions of the numerical Q-ball radius, $r_{X}$ and $r_{Z}$, in order to compare the results. The first, $r_{X}$ (referred to from here when discussing the quantities measured in this definition as the "$X$ point"), is defined as the distance from the centre of the Q-ball at which the field has decreased to $1 \%$ of its initial value $\phi_{0}$. This is so that we can examine the properties using a concrete definition of what the Q-ball radius is. The second, $r_{Z}$, (henceforth referred to as the "$Z$ point"), is the point at which the code used to determine the exact numerical solution to the Q-ball equation cuts off, having successfully found a Q-ball solution. Since the numerical solution is never exact, it will stop its asymptote towards zero and the field will begin to increase again and break away from the Q-ball solution at a a particular value of $r$. This is $r_{Z}$ and it denotes the maximum value of $r$ for which the Q-ball solution is valid. This is somewhat less well defined than the $X$ point, since the program will run to a different end point for each Q-ball. However, it is useful to include a measure of the properties at this point since this is the largest value of $r$ to which the Q-ball equation is integrated, and so this will give the best estimate of the charge and energy of the Q-balls. All quantities listed with a subscript $X$ are determined using the $X$ point, and all those with a $Z$ are determined using the $Z$ point.

In Figures $5$-$8$ we illustrate two examples of Q-ball solutions, calculated  both numerically and using an analytical approximation, \eq{e100}, which is derived in the next section. The first example shown is a Q-ball with a large starting field value $\phi_{0} \sim 10^{16}\GeV$, and the second is a Q-ball with $\phi_{0} \sim 10^{13}\GeV$, with $\omega$ close to the upper bound of compatibility with the existence of Q-balls, $\omega = m$. The inflaton self-coupling is $\lambda = 0.1$ throughout and the corresponding non-minimal coupling is $\xi = 1.2163 \times 10^{9}$.

\begin{figure}[H]
\begin{center}
\includegraphics[clip = true, width=0.75\textwidth, angle = 360]{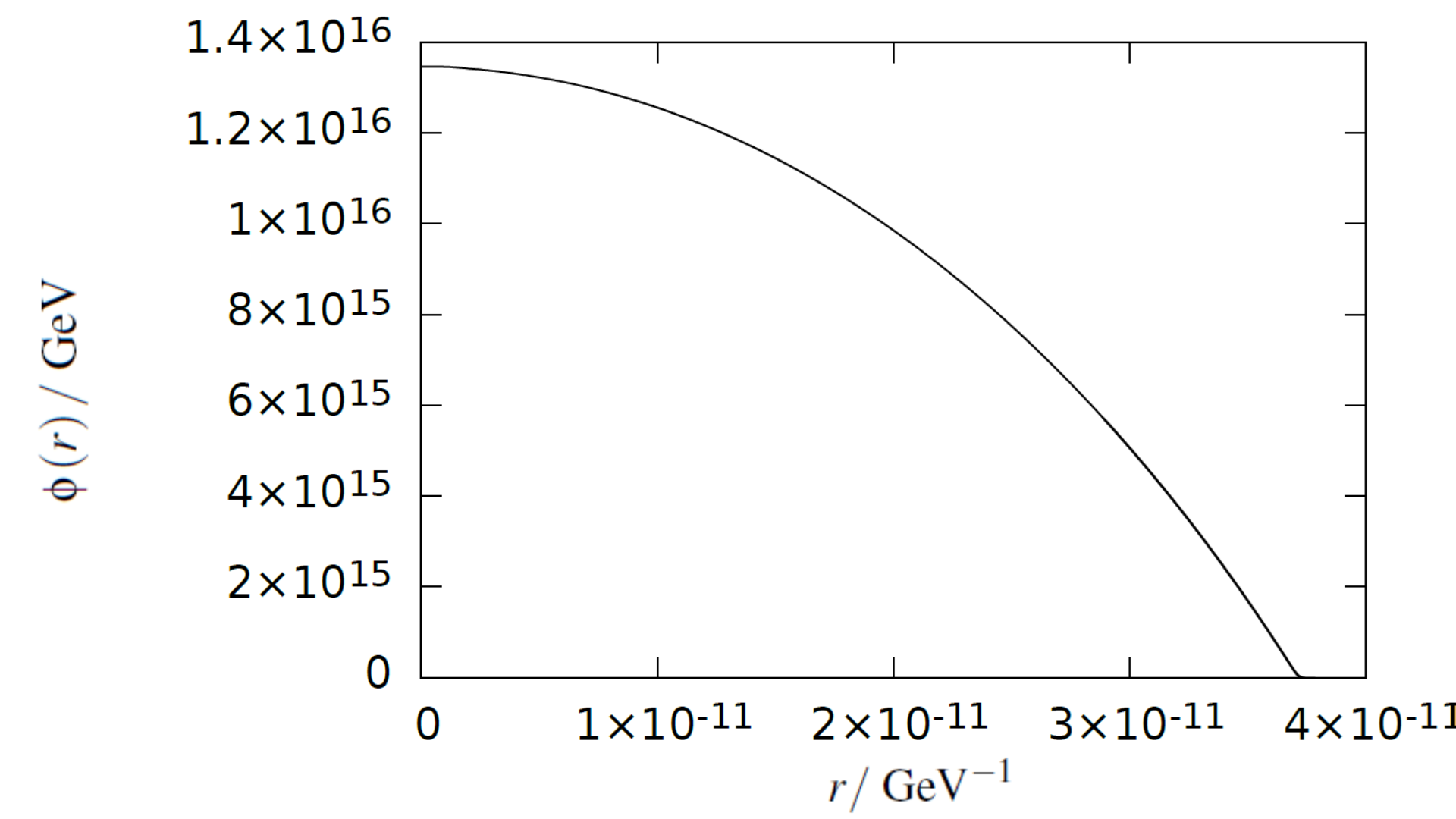}
\caption{Plot illustrating a Q-ball with $\omega = 0.709\omega_{c} $, $m=0.9\omega_{c}$ and $\phi_{0} = 1.3464098 \times 10^{16} \GeV$, showing the Q-ball profile, $\phi \left(r\right)$, from the numerical calculation.} 
\label{fig5}
\end{center}
\end{figure} 

\vspace{1cm}

\begin{figure}[H]
\begin{center}
\includegraphics[clip = true, width=0.75\textwidth, angle = 360]{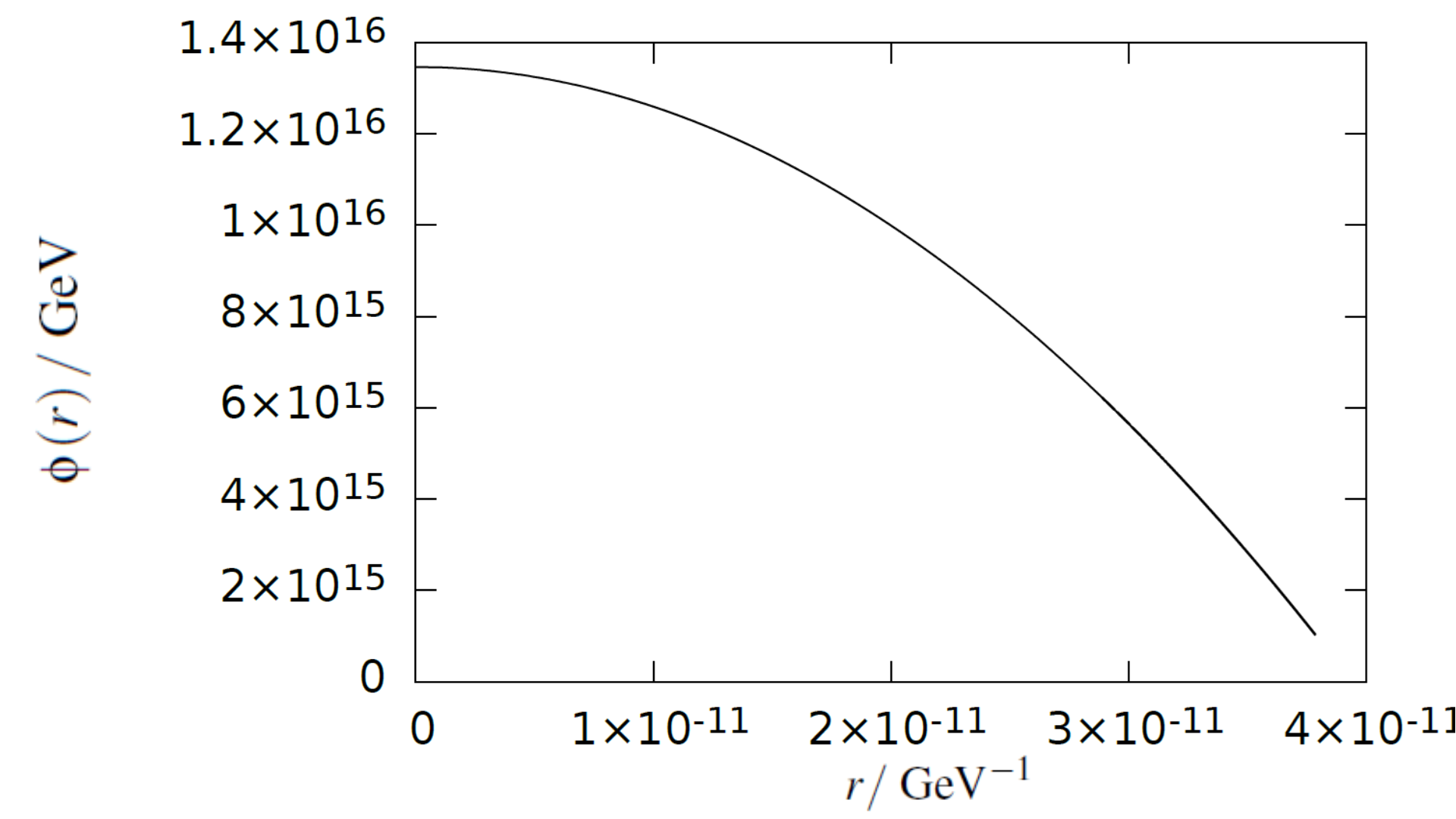}
\caption{Plot illustrating a Q-ball with $\omega = 0.709\omega_{c} $, $m=0.9\omega_{c}$ and $\phi_{0} = 1.3464098 \times 10^{16} \GeV$, showing the Q-ball profile, $\phi \left(r\right)$, from the analytical approximation \eq{e104}.} 
\label{fig6}
\end{center}
\end{figure}

\begin{figure}[H]
\begin{center}
\includegraphics[clip = true, width=0.75\textwidth, angle = 360]{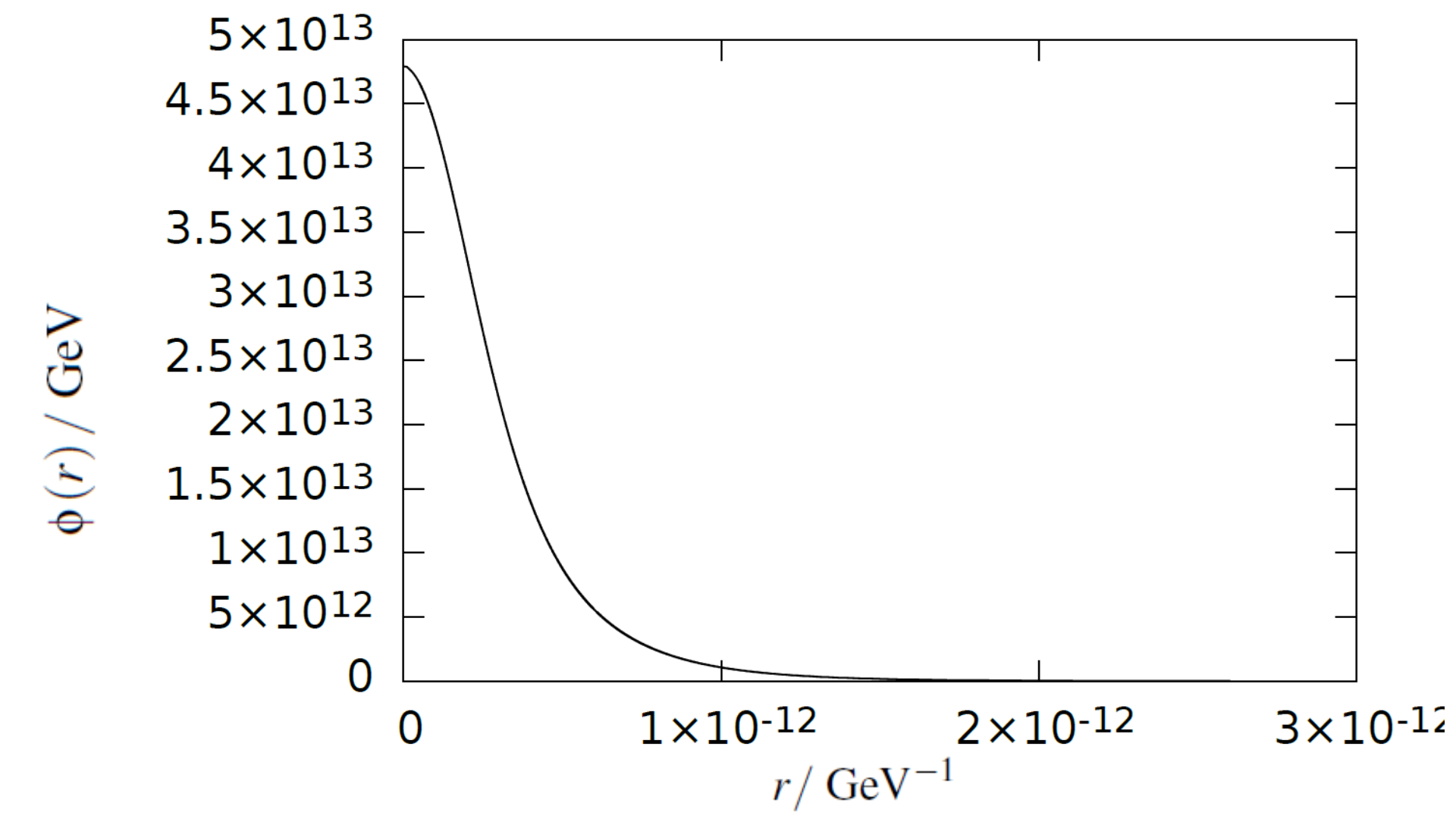}
\caption{Plot illustrating a Q-ball with $\omega = 0.89\omega_{c}$, $m=0.9\omega_{c}$ and $\phi_{0} = 4.7918\times 10^{13} \GeV$, showing the Q-ball profile for the numerical calculation. } 
\label{fig7}
\end{center}
\end{figure}

\begin{figure}[H]
\begin{center}
\includegraphics[clip = true, width=0.75\textwidth, angle = 360]{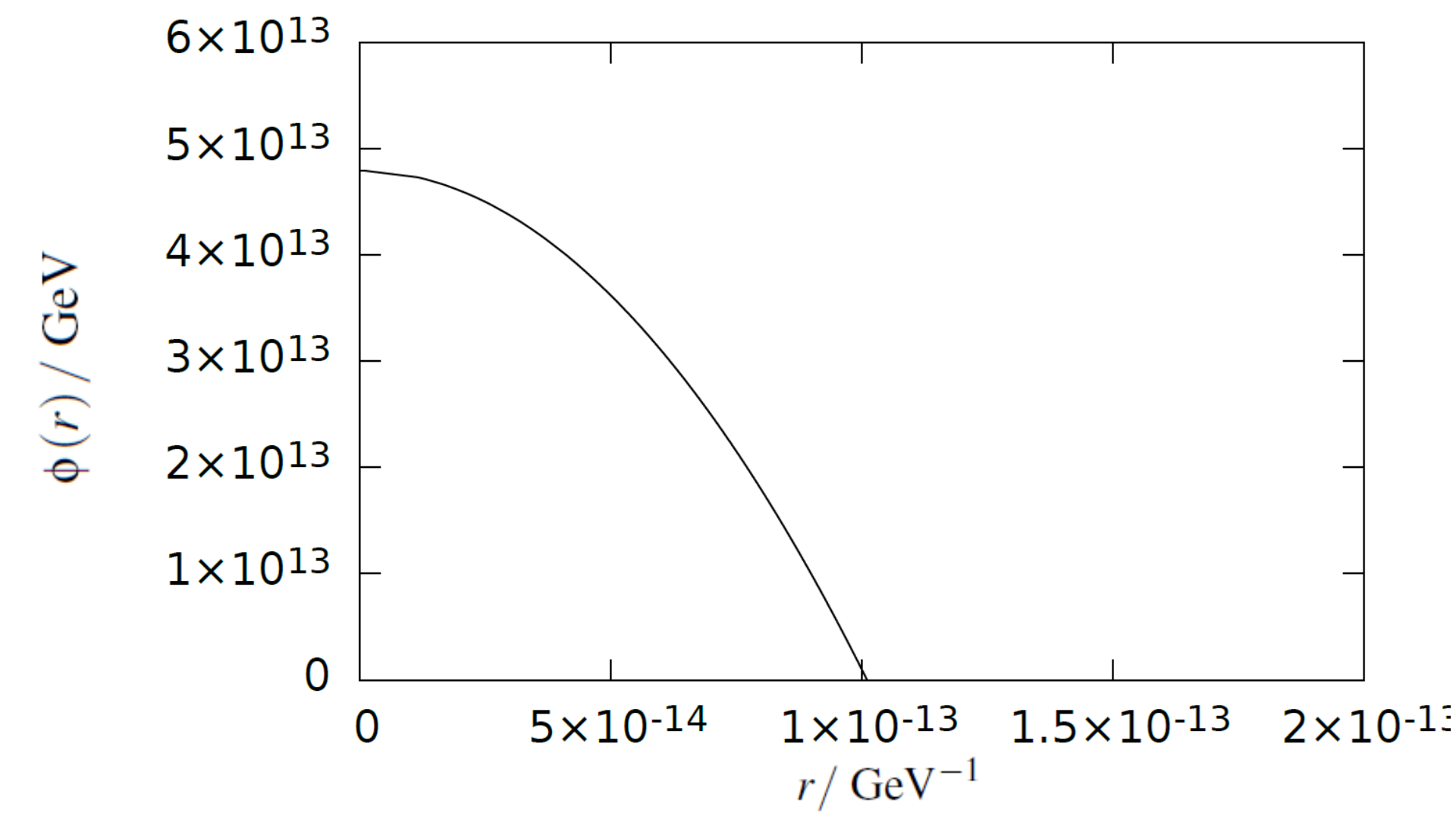}
\caption{Plot illustrating a Q-ball with $\omega = 0.89\omega_{c}$, $m=0.9\omega_{c}$ and $\phi_{0} = 4.7918\times 10^{13} \GeV$, showing the Q-ball profile for the analytical approximation \eq{e104}. } 
\label{fig8}
\end{center}
\end{figure}

Tables 1 and 2 show the numerical properties of the Q-balls from integrating to $r_{X}$ and $r_{Z}$, respectively. The numbers shown in the tables are the $\omega$ parameter, the initial $r = 0$ value of the field profile for the Q-ball solution $\phi_{0}$, the value of $r_{X}$ or $r_{Z}$, the value of the field at the defined radius of the Q-ball, the total energy, $E$, charge, $Q$, and the energy-charge ratio, $E/Q$.

We also check the stability of the Q-ball solutions we have obtained. To do this we will introduce two quantities which will be useful in this analysis, $\Delta_{\omega}$ and $\Delta_{m}$. 
$\Delta_{\omega}$ is defined by
\be{ e92  } 
\Delta_{\omega} = \frac{1}{\omega}\left(\frac{E}{Q} - \omega \right).
\ee
This quantity allows us to determine the extent to which the $\omega$ parameter for each of the Q-ball solutions is equal to the energy-charge ratio.
 
$\Delta_{m}$ is defined by
\be{ e93  } 
\Delta_{m} = \frac{1}{m}\left(\frac{E}{Q} - m \right).
\ee
This quantity provides a measure of absolute stability. As we discussed in Section $3$, absolute stability is determined by whether or not a given Q-ball is a lower energy configuration than the equivalent number of free scalars i.e. $E < mQ$. If $\Delta_{m}$ is less than zero for a given Q-ball then this Q-ball is numerically absolutely stable. In addition, the magnitude of $\Delta_{m}$ is a measure of the binding energy of the scalars in the Q-ball relative to their mass.

\begin{table}[H]
\begin{center}
\begin{tabular}{| c | c | c | c | c | c | c | c | c |}
\hline
$\omega / \omega_{c}$ & $\phi_{0} / \GeV$ & $\phi_{Z}/\GeV$ & $r_{Z}/\GeV^{-1}$ & $Q_{Z}$ & $E_{Z}/\GeV$ & $(E/Q)_{Z}/\GeV$ & $\Delta_{\omega, Z}$ & $ \Delta_{m, Z}$ \\
\hline
$0.707155$ & $3.2217991 \times 10^{17} $ & $1.61 \times 10^{10}$ & $8.94 \times 10^{-10}$ & $2.17 \times 10^{14}$ & $3.34 \times 10^{27}$ & $1.54 \times 10^{13}$ & $-2.79 \times 10^{-4}$ & $-0.21$ \\
\hline
$0.709$ & $1.3464098 \times 10^{16}$ & $1.63 \times 10^{10}$ & $3.79 \times 10^{-11}$ & $1.56 \times 10^{10}$ & $2.40 \times 10^{23}$ & $1.54 \times 10^{13}$ & $9.71 \times 10^{-4}$ & $-0.21$\\
\hline
$0.71$ & $8.855792 \times 10^{15}$ & $2.05 \times 10^{10}$ & $2.50 \times 10^{-11}$ & $4.37 \times 10^{9}$ & $6.76 \times 10^{22}$ & $1.55 \times 10^{13}$ & $1.64 \times 10^{-3}$ & $-0.21$ \\
\hline
$0.72$ & $1.960795 \times 10^{15}$ & $9.76 \times 10^{9}$ & $6.09 \times 10^{-12}$ &  $4.43 \times 10^{7}$ & $6.99 \times 10^{20}$ & $1.58 \times 10^{13}$ & $7.71 \times 10^{-3}$ & $-0.19$ \\
\hline
$0.73$ & $1.090258 \times 10^{15}$ & $6.60 \times 10^{9}$ & $3.73 \times 10^{-12}$ &  $7.09 \times 10^{6}$ & $1.14 \times 10^{20}$ & $1.61 \times 10^{13}$ & $1.30 \times 10^{-2}$ & $-0.18$ \\
\hline
$0.74$ & $7.45339 \times 10^{14}$ & $1.61 \times 10^{10}$ & $2.72 \times 10^{-12}$ &  $2.16 \times 10^{6}$ & $3.54 \times 10^{19}$ & $1.64 \times 10^{13}$ & $1.76 \times 10^{-2}$ & $-0.16$ \\
\hline
$0.75$ & $5.61953 \times 10^{14}$ & $1.14 \times 10^{10}$ & $2.28 \times 10^{-12}$ & $8.85 \times 10^{5}$ & $1.48 \times 10^{19}$ & $1.67 \times 10^{13}$ & $2.17 \times 10^{-2}$ & $-0.15$ \\
\hline
$0.80$ & $2.29632 \times 10^{14}$ & $1.96 \times 10^{10}$ & $1.47 \times 10^{-12}$ & $5.60 \times 10^{4}$ & $1.01 \times 10^{18}$ & $1.80 \times 10^{13}$ & $3.51 \times 10^{-2}$ & $-8.00 \times 10^{-2}$ \\
\hline
$0.85$ & $1.16877 \times 10^{14}$ & $1.57 \times 10^{10}$ & $1.49 \times 10^{-12}$ & $1.01 \times 10^{4}$ & $1.93 \times 10^{17}$ & $1.91 \times 10^{13}$ & $3.51 \times 10^{-2}$ & $-2.24 \times 10^{-2}$ \\
\hline
$0.89$ & $4.7918 \times 10^{13}$ & $8.82 \times 10^{9}$ & $2.70 \times 10^{-12}$ &  $3.98 \times 10^{3}$ & $7.82 \times 10^{16}$ & $1.97 \times 10^{13}$ & $1.50 \times 10^{-2}$ & $3.74 \times 10^{-3}$ \\
\hline
\end{tabular}
\caption{Table illustrating Q-ball properties for the $m=0.9\omega_{c}$ Q-balls using the $Z$ point.  $\phi_{0}$ is the value of the scalar field at $r = 0$. $\phi_{Z}$, $E_{Z}$, $Q_{Z}$, $(E/Q)_{Z}$, $\Delta_{\omega, Z}$ and $\Delta_{m, Z}$ are calculated numerically at the point $r_{Z}$, which we define as the point where the code cuts off the Q-ball solution.}
\end{center}
\end{table}

\begin{table}[H]
\begin{center}
\begin{tabular}{| c | c | c | c | c | c | c | c | c |}
\hline
$\omega / \omega_{c}$ & $\phi_{0}/\GeV$ & $\phi_{X}/\GeV$ & $r_{X}/\GeV^{-1}$ & $Q_{X}$ & $E_{X}/\GeV$ & $(E/Q)_{X}/\GeV$ & $\Delta_{\omega, X}$ & $\Delta_{m, X}$\\
\hline
$0.707155$ & $3.2217991 \times 10^{17} $ & $3.22 \times 10^{15}$ & $8.89 \times 10^{-10}$ & $2.15 \times 10^{14}$ & $3.30 \times 10^{27}$ & $1.54 \times 10^{13}$ & $-4.47 \times 10^{-4}$ & $-0.22$ \\
\hline
$0.709$ & $1.3464098 \times 10^{16}$ & $1.35 \times 10^{14}$ & $3.70 \times 10^{-11}$ & $1.54 \times 10^{10}$ & $2.38 \times 10^{23}$ & $1.54 \times 10^{13}$ & $ -1.27 \times 10^{-3}$ & $-0.21$ \\
\hline
$0.71$ & $8.855792 \times 10^{15}$ & $8.86 \times 10^{13}$ & $2.43 \times 10^{-11}$ & $4.34 \times 10^{9}$ & $6.70 \times 10^{22}$ & $1.54 \times 10^{13}$ & $-7.78 \times 10^{-4}$ & $-0.21$ \\
\hline
$0.72$ & $1.960795 \times 10^{15}$ & $1.96 \times 10^{13}$ & $5.38 \times 10^{-12}$ & $4.42 \times 10^{7}$ & $6.97 \times 10^{20}$ & $1.58 \times 10^{13}$ & $6.58 \times 10^{-3}$ & $-0.19$\\
\hline
$0.73$ & $1.090258 \times 10^{15}$ & $1.09 \times 10^{13}$ & $3.03 \times 10^{-12}$ & $7.08 \times 10^{6}$ & $1.14 \times 10^{20}$ & $1.61 \times 10^{13}$ & $1.23 \times 10^{-2}$ & $-0.18$\\
\hline
$0.74$ & $7.45339 \times 10^{14}$ & $7.45 \times 10^{12}$ & $2.13 \times 10^{-12}$ & $2.16 \times 10^{6}$ & $3.54 \times 10^{19}$ & $1.64 \times 10^{13}$ & $1.71 \times 10^{-2}$ & $-0.16$ \\
\hline
$0.75$ & $5.61953 \times 10^{14}$ & $5.62 \times 10^{12}$ & $1.66 \times 10^{-12}$ & $8.85 \times 10^{5}$ & $1.48 \times 10^{19}$ & $1.67 \times 10^{13}$ & $2.13 \times 10^{-2}$ & $-0.15$\\
\hline
$0.80$ & $2.29632 \times 10^{14}$ & $2.30 \times 10^{12}$ & $8.97 \times 10^{-13}$ & $5.60 \times 10^{4}$ & $1.01 \times 10^{18}$ & $1.80 \times 10^{13}$ & $3.48 \times 10^{-2}$ & $-8.02 \times 10^{-2}$ \\
\hline
$0.85$ & $1.16877 \times 10^{14}$ & $1.17 \times 10^{12}$ & $7.84 \times 10^{-13}$ & $1.01 \times 10^{4}$ & $1.93 \times 10^{17}$ & $1.91 \times 10^{13}$ & $3.49 \times 10^{-2}$ & $-2.26 \times 10^{-2}$ \\
\hline
$0.89$ & $4.7918 \times 10^{13}$ & $4.79 \times 10^{11}$ & $1.21 \times 10^{-12}$ & $3.96 \times 10^{3}$ & $7.79 \times 10^{16}$ & $1.97 \times 10^{13}$ & $1.50 \times 10^{-2}$ & $3.69 \times 10^{-3}$ \\
\hline
\end{tabular}
\caption{Table illustrating Q-ball properties for the $m=0.9\omega_{c}$ Q-balls using the $X$ point. $\phi_{0}$ is the value of the scalar field at $r = 0$. $\phi_{X}$, $E_{X}$, $Q_{X}$, $(E/Q)_{X}$, $\Delta_{\omega, X}$ and $\Delta_{m, X}$ are calculated numerically at the point $r_{X}$, which we define as the point where the value of the field drops to $0.01\phi_{0}$.}
\end{center}
\end{table}

We note that the energy-charge ratio increases by less than an order of magnitude across the data from the largest $\phi_{0}$ Q-ball to the smallest. The data also shows that there is no significant difference in magnitude in the energy and charge calculated between the $X$ and $Z$ points, and $E$, $Q$ and $E/Q$ at the $X$ point are comparable to their $Z$ point counterparts. 

The larger $\phi_{0}$ Q-balls have larger radii and, for $\phi_{0}$ large compared to $M_{Pl}/\sqrt{\xi}  \; (= 6.9 \times 10^{13} \GeV)$, the radius increases linearly with $\phi_{0}$ to a good approximation. This is in agreement with the analytical approximation, which is discussed in the next section.  The energy and charge of the Q-balls also increase with increasing $\phi_{0}$. This is expected as the larger the radius of the Q-ball, the greater the number of component scalars it will be composed from, and it will therefore carry a greater charge and energy.  Figures $9$ and $10$ illustrate the relationship between the logarithm of the energy and charge and the logarithm of $\phi_{0}$. These show the proportionality of $E$ and $Q$ to $\phi_{0}^{3}$ for the larger $\phi_{0}$ Q-balls, with the proportionality becoming becoming weaker for the smaller $\phi_{0}$ Q-balls. This relationship is predicted by the analytical approximation derived in the next section, and it therefore provides a useful test of the validity of the numerical Q-ball solutions.

\begin{figure}[H]
\begin{center}
\includegraphics[clip = true, width=0.75\textwidth, angle = 360]{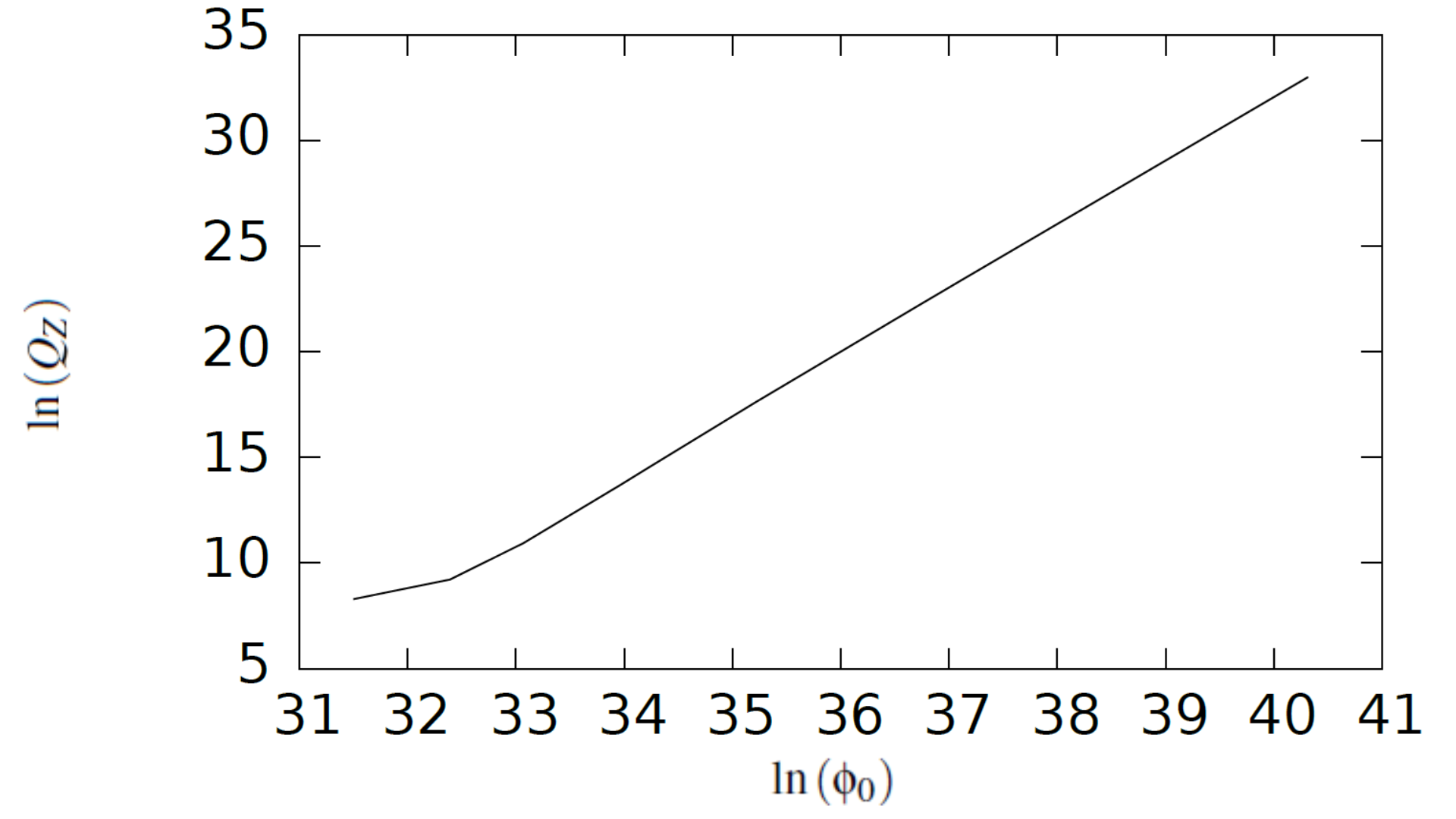}
\caption{Q-ball charge, $Q_{z}$, vs. $\phi_{0}$ for the $m=0.9\omega_{c}$ Q-balls.}
\label{fig9}
\end{center}
\end{figure} 

\begin{figure}[H]
\begin{center}
\includegraphics[clip = true, width=0.75\textwidth, angle = 360]{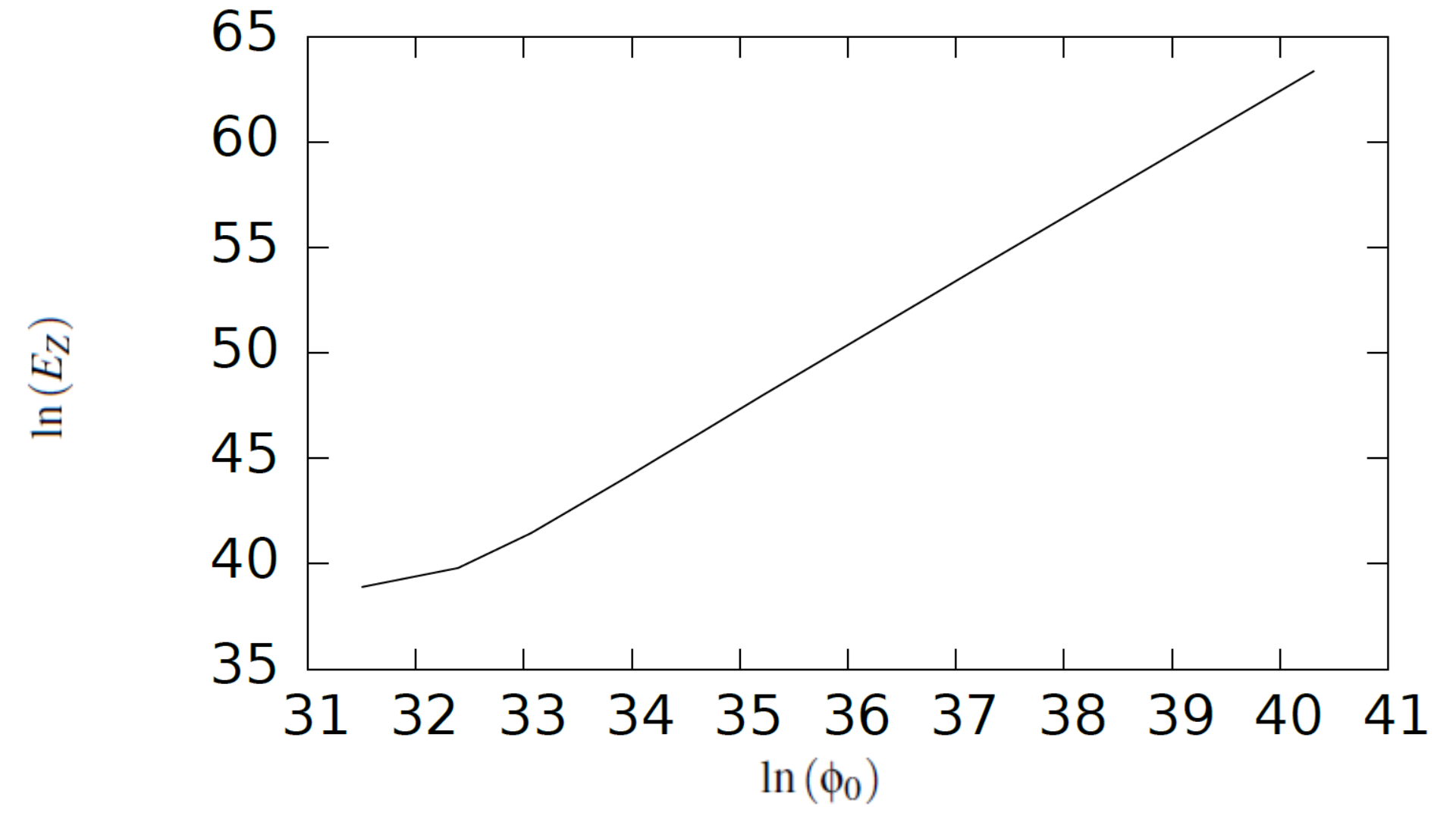}
\caption{Q-ball energy, $E_{z}$, vs. $\phi_{0}$ for the $m=0.9\omega_{c}$ Q-balls.}
\label{fig10}
\end{center}
\end{figure}

\noindent In general, $\left| \Delta_{\omega} \right| \sim 10^{-4} - 10^{-2}$. This shows that for all of the Q-balls in the tables $\omega$ is close to the value of the energy-charge ratio, and therefore to a good approximation the $\omega$ parameter represents the chemical potential of the Q-balls in this model. We also note that $ \mid \Delta_{\omega} \mid $ gets larger the smaller the Q-balls get, and we see in Figure $11$ that $ \Delta_{\omega} $ overall increases as Q-ball radius decreases. This is consistent with what we have seen so far, in that the behaviour of the solutions is somewhat less regular for the smaller $\phi_{0}$ Q-balls than for the larger $\phi_{0}$ Q-balls. This is likely to be due to the approach of $\phi_{0}$ to the value $M_{Pl}/\sqrt{\xi}$ at which the potential deviates from the plateau and where the Q-balls become unstable.

\begin{figure}[H]
\begin{center}
\includegraphics[clip = true, width=0.75\textwidth, angle = 360]{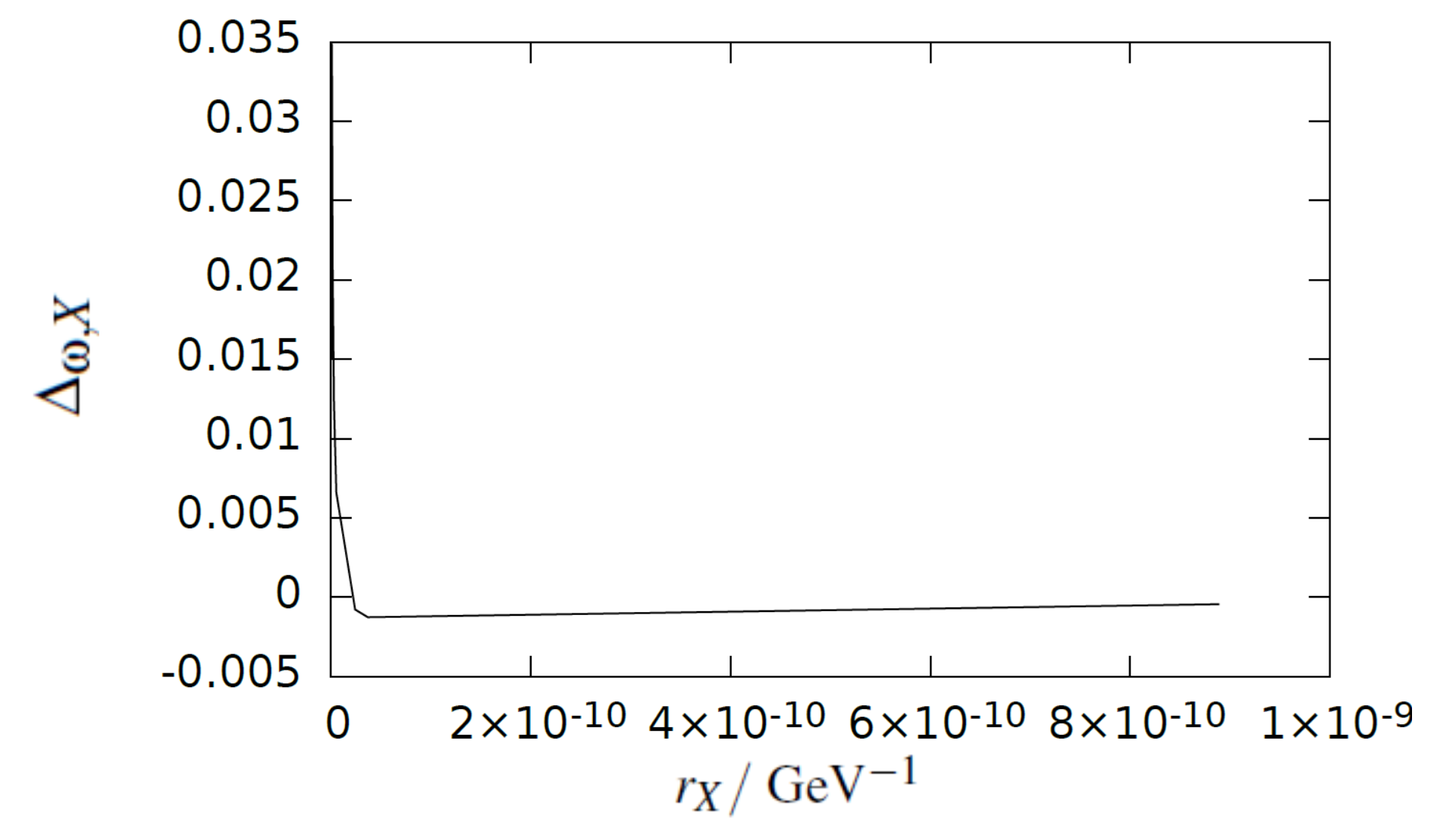}
\caption{$\Delta_{\omega,X}$ vs $r_{X}$ for the $m=0.9\omega_{c}$ Q-balls. } 
\label{fig11}
\end{center}
\end{figure}

We find that $\Delta_{m}$ is negative for all cases with the exception of $\Delta_{m}$ for the $\omega = 0.89\omega_{c}$ Q-ball, as illustrated in Figure $12$. This confirms that all of the Q-balls are absolutely stable except for the $\omega = 0.89\omega_{c}$ case. The magnitude of $\Delta_{m}$ for $\omega = 0.89\omega_{c}$ is small, therefore it is possible that $\Delta_{m}$ is actually negative but smaller than the level of numerical errors in the Q-ball calculation. Alternatively, there may exist metastable Q-ball solutions with $\omega < m$ but $E/Q > m$. We will show in Section 7 that as $\phi_{0}$ approaches $M_{Pl}/\sqrt{\xi}$, $E/Q$ becomes significantly larger than $\omega$ and hence can approach $m$ even if $\omega < m$.

\begin{figure}[H]
\begin{center}
\includegraphics[clip = true, width=0.75\textwidth, angle = 360]{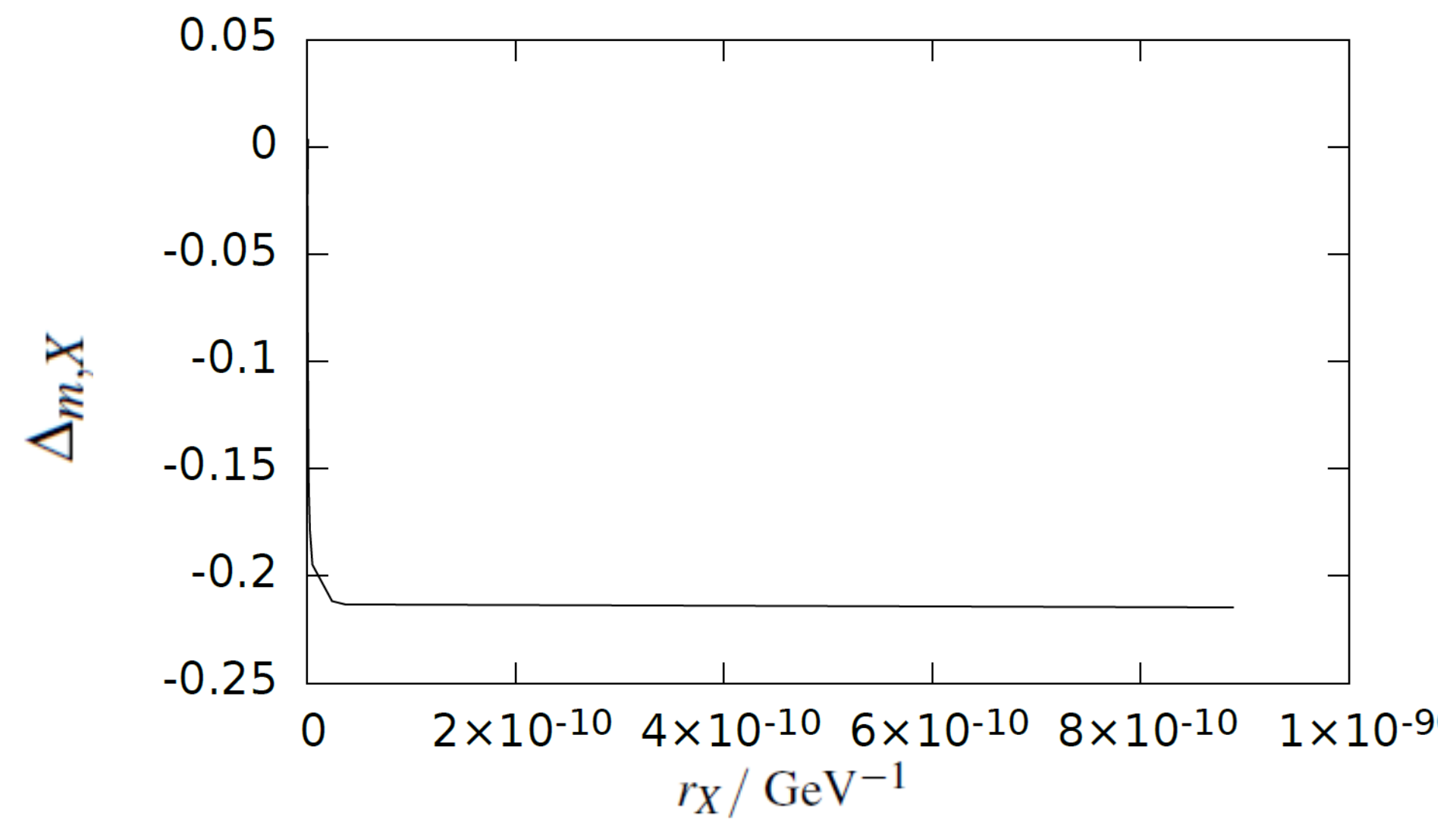}
\caption{$\Delta_{m,X}$ vs $r_{X}$ for the $m=0.9\omega_{c}$ Q-balls. } 
\label{fig12}
\end{center}
\end{figure}

From the point of view of Palatini inflation cosmology, it is significant that we can obtain Q-balls with $\phi_{0} \gtrsim 10^{17}\GeV$, and that these are absolutely stable. This means that Q-balls with the magnitude of $\phi$ that we would naturally expect from inflaton condensate fragmentation following tachyonic preheating \cite{rubio19} at the end of non-minimally coupled Palatini inflation, $\phi \sim 10^{17}-10^{18}\GeV$ (Appendix B), are stable. Therefore, should such Q-balls form, they may have a non-trivial and perhaps measureable effect on the post-inflationary evolution of this model.

\section{An Analytical Approximation to the Q-ball Solution}

In the previous section we demonstrated the existence of Q-ball solutions for values of the inflaton mass squared within the Q-ball window for solutions consistent with Palatini inflation, and we showed that it is possible for the field inside the Q-balls to be as large as the typical inflaton field following the end of Palatini inflation, $\phi_{0} \sim 10^{17}-10^{18} \GeV$.
In order to understand the properties of these Q-balls, in this section we discuss an analytical approximation to the Q-ball solution, and we calculate the energy and charge of the Q-balls in this approximation\footnote{Previous studies utilising analytic approximations as an estimate of Q-ball properties have been carried out in e.g. \cite{multamaki99}, \cite{pc01}, \cite{ioannidou03} and \cite{tsumagari08}.}.

Assuming that we are high on the inflationary plateau, the right hand side of the Q-ball equation \eq{e63} can be approximated as

\be{ e94 } 
\Omega^{2}\frac{\partial V_{\omega}}{\partial \phi} \approx - \frac{\gamma}{\phi} ~,
\ee

\noindent where 

\be{ e95 } 
\gamma = \frac{M_{pl}^{2}}{\xi}\left( m^{2} + \omega^{2} - \frac{\lambda M_{pl}^{2}}{\xi}\right) = \frac{M_{pl}^{2}}{\xi}\left( m^{2} + \omega^{2} - \omega_{c}^{2} \right),
\ee
We also assume that, for $\phi >> M_{pl}/\sqrt{\xi}$,
\be{ e96 } 
\left| \frac{\partial^{2}\phi}{\partial r^{2}} + \frac{2}{r}\frac{\partial \phi}{\partial r} \right| >> K\left(\phi \right) \left(\frac{\partial \phi}{\partial r}\right)^{2} 
\ee

\noindent is satisfied, where $K\left(\phi \right) \sim 1/\phi$ in the plateau limit. This assumption is satisfied for the resulting analytical solution. This leaves the approximate Q-ball equation as

\be{e97} 
\frac{\partial^{2}\phi}{\partial r^{2}} + \frac{2}{r}\frac{\partial \phi}{\partial r} \approx - \frac{\gamma}{\phi}.
\ee

\noindent where $\gamma > 0$ for Q-ball solutions to exist.

\noindent For a sufficiently small $r$, $\phi$ will not deviate from its initial value $\phi_{0}$ too much.  In general, we have $\partial \phi/\partial r \rightarrow 0$ as $r \rightarrow 0$.
We therefore set $\phi = \phi_{0}$ on the right hand side of \eq{e97} and make the following ansatz for the solution of \eq{e97}
\be{ e98 } 
\phi \left( r \right) = \phi_{0} - A r^{2}
\ee
\noindent where $A$ is some coefficient to be determined.
Substituting into \eq{e97}, we obtain
\be{ e99 } 
A = \frac{\gamma}{6 \phi_{0}}.
\ee

\noindent So at small r, we can approximate the field profile to be

\be{e100} 
\phi \left( r \right) = \phi_{0} -  \frac{\gamma}{6 \phi_{0}}r^{2}.
\ee

\noindent The assumption that $\phi \thickapprox \phi_{0}$ holds reasonably well if

\be{ e101 } 
r^{2} < \frac{r_{Q}^{2}}{4} 
\ee

where
\be{e102}
r_{Q} = \frac{\sqrt{6} \phi_{0}}{\sqrt{\gamma}}.
\ee
\noindent To a good approximation, we can therefore say that the Q-ball solution at $r < r_{Q}/2$ is given 
by

\be{e103} 
\phi \left( r \right) = \phi_{0}\left[ 1 - \left(\frac{r}{r_{Q}}\right)^{2} \right]
\ee

 The analytical estimate of the Q-ball energy and charge provide useful lower bounds on the values of $E$ and $Q$ for comparison with the numerical solution, and allow us to understand their parametric dependence on the radius of the Q-ball. In the plateau limit, assuming $\phi \thickapprox \phi_{0}$, such that $\phi_{0}^{2} \gg M_{Pl}^{2}/\xi$ and $\Omega^{2} \approx \xi \phi_{0}^{2}/M_{Pl}^{2}$, we find that the energy $E$ of the approximate solution is

\be{e104a} E = \int_{0}^{r_{Q}/2} dr \;4 \pi r ^{2} \rho_{E} = \int_{0}^{r_{Q}/2} dr \;4 \pi r ^{2} \left[ \frac{1}{2 \Omega^{2}} \left(\frac{\partial \phi}{\partial r} \right)^{2} + \frac{\omega^{2} \phi^{2}}{2 \Omega^{2}} + \frac{V(\phi)}{\Omega^{4}} \right]  ~\ee

\be{e104} 
\Rightarrow E \approx \frac{\pi M_{pl}^{2}}{12 \xi}\left[ \omega^{2} + \frac{\lambda M_{pl}^{2}}{2\xi}\right] r_{Q}^{3} \equiv  \frac{\pi M_{pl}^{2}}{12 \xi}\left[ \omega^{2} + \frac{\omega_{c}^{2}}{2}\right] r_{Q}^{3} ~. 
\ee

\noindent Note that this expression integrates the energy density for $r < r_{Q}/2$ and is therefore a lower bound on the true Q-ball energy. Similarly, the charge of the approximate solution is 

\be{e105a} Q = \int_{0}^{r_{Q}/2} dr \;4 \pi r ^{2} \rho_{Q} = \int_{0}^{r_{Q}/2} dr \;4 \pi r ^{2} \frac{\omega \phi^{2}}{\Omega^{2}}   ~\ee

\be{e105} 
\Rightarrow Q \approx \frac{\pi \omega M_{pl}^{2}}{6 \xi} r_{Q}^{3}
~.\ee 

\noindent  Therefore both $E$ and $Q$ are proportional to $r_{Q}^{3}$ and so to the volume of the Q-ball. Examining the expression for $r_{Q}$, \eq{e102}, and assuming that $\omega$ is insensitive\footnote{We expect that $\omega$ will equal $m$ up to a small suppression of the binding energy of the Q-balls. This is consistent with the numerical Q-ball solutions.} to $\phi_{0}$, we see that $r_{Q}$ is proportional to $\phi_{0}$, therefore we can translate this to a proportionality of $E$ and $Q$ to $\phi_{0}^{3}$. These relationships were confirmed numerically in the previous section, where they we found to be very accurate for the large $\phi_{0}$ Q-balls.

The analytical approximation prediction for $E/Q$ is  
\be{e106}   \frac{E}{Q} = \frac{1}{2} \left[ \omega + \frac{\omega_{c}^{2}}{2 \omega} \right]    ~.\ee

\noindent This does not reproduce the numerical result $E/Q = \omega$, which is observed to be true up to a very small deviation when $\phi_{0} \gg M_{Pl}/\sqrt{\xi}$, although the magnitude from \eq{e106} is correct. However, this is not surprising, as we only integrate $E$ and $Q$ to $r = r_{Q}/2$. In the next section we will derive an exact expression for $E/Q$ in terms of $\phi(r)$ that will allow us to use the analytical approximation to accurately model the behaviour of $E/Q$ as a function of $\phi_{0}$.

In Figures 6 and 8 we show the analytical approximate solution for $\phi$, \eq{e100}, as a function of $r$. Comparing with the numerical solutions shown in Figures 5 and 7, we see that the analytical approximation closely follows the numerical solution for small $r$ and is a good fit to the exact solution for $r < r_{Q}$ in the case where $\phi_{0} >> M_{pl}/\sqrt{\xi}$ , but it becomes a poor match when $\phi_{0}$ approaches $M_{Pl}/\sqrt{\xi}$. Therefore $r_Q$ gives a good estimate of the Q-ball radius when $phi_0$ is large. This shows that it is reasonable to use the analytical approximation to estimate of the properties of the larger $\phi_{0}$ Q-balls.

\begin{table}[H]
\begin{center}
\begin{tabular}{| c | c | c | c | c |}
\hline
$\omega / \omega_{c}$ & $\phi_{0}/\GeV$ & $r_{Q}/\GeV^{-1}$ & $Q$  & $E/\GeV$ \\
\hline
$0.707155$ & $3.2217991 \times 10^{17} $ & $9.46 \times 10^{-10}$ & $3.23 \times 10^{13}$ & $4.97 \times 10^{26}$\\
\hline
$0.709$ & $1.3464098 \times 10^{16}$ & $3.94 \times 10^{-11}$ & $2.34 \times 10^{9}$ & $3.60 \times 10^{22}$\\
\hline
$0.71$ & $8.855792 \times 10^{15}$ & $2.58 \times 10^{-11}$ & $6.58 \times 10^{8}$ & $1.01 \times 10^{22}$\\
\hline
$0.72$ & $1.960795\times 10^{15}$ & $5.60 \times 10^{-12}$ & $6.82 \times 10^{6}$ & $1.05 \times 10^{20}$ \\
\hline
$0.73$ & $1.090258 \times 10^{15}$ & $3.05 \times 10^{-12}$ & $1.12 \times 10^{6}$ & $1.72 \times 10^{19}$ \\
\hline
$0.74$ & $7.45339 \times 10^{14}$ & $2.04 \times 10^{-12}$ & $3.39 \times 10^{5}$ & $5.22 \times 10^{18}$ \\
\hline
$0.75$ & $5.61953 \times 10^{14}$ & $1.51 \times 10^{-12}$ & $1.39 \times 10^{5}$ & $2.15 \times 10^{18}$ \\
\hline
$0.80$ & $2.29632 \times 10^{14}$ & $5.60 \times 10^{-13}$ & $7.58 \times 10^{3}$ & $1.18 \times 10^{17}$ \\
\hline
$0.85$ & $1.16877 \times 10^{14}$ & $2.62 \times 10^{-13}$ & $825$ & $1.29 \times 10^{16}$ \\
\hline
$0.89$ & $4.7918 \times 10^{13}$ & $1.01 \times 10^{-13}$ & $50$ & $7.82 \times 10^{14}$ \\
\hline
\end{tabular}
\caption{Table illustrating $r_{Q}$ and the analytical approximation values of $E$ and $Q$ as a function of $\phi_{0}$ and $\omega$ for the $m=0.9\omega_{c}$ Q-balls.}
\end{center}
\end{table}

\noindent Table $3$ shows the estimates of the energy and charge calculated using the analytical approximation expressions \eq{e104} and \eq{e105}. Comparing these data to those in Tables 1 and 2 we can see that the analytical estimates consistently underestimate the energy and charge compared to the numerical results by about an order of magnitude for most of the Q-balls. Since the analytical estimates of $E$ and $Q$ are valid up to about $r = r_{Q}/2$, it makes sense that this would provide an underestimate of the energy and charge by a factor of around $(1/2)^{3} \sim 0.1$. For the smallest $\phi_{0}$ Q-balls the underestimate is much greater, and the analytical approximation does not provide a reliable estimate of the energy and charge. This is consistent with the poor fit of the analytical Q-ball profile $\phi(r)$ to the numerical profile when $\phi_{0}$ is close to the plateau limit. The analytical approximation is therefore a good approximation for the larger $\phi_{0}$ Q-balls, and can be used to deduce their parameter dependence. 

As an example, we can use the analytical approximation to determine how the Q-ball solutions depend on $\lambda$. For a fixed value of $m/\omega_{c}$, we have $\omega \approx m$ and therefore both $m$ and $\omega$ are proportional to $\omega_{c}$. Thus $\gamma \propto \omega_{c}^{2}/\xi$. From \eq{e34}, we find that $\xi \propto \lambda$ once it is normalised to the observed power spectrum. Therefore $\gamma \propto 1/\lambda$. From \eq{e102} we then find that 
\be{e1d}  r_{Q} \propto \sqrt{\lambda} \phi_{0}   ~.\ee 
From \eq{e104}, with $\omega^{2} \propto \omega_{c}^{2}$, we find that $E \propto \omega_{c}^{2} r_{Q}^{3}/\xi$. Since $\xi \propto \lambda$, we have $\omega_{c} \propto \sqrt{\lambda}/\sqrt{\xi}$. Therefore, since $r_{Q} \propto \sqrt{\lambda} \phi_{0}$, we obtain 
\be{e2d} E \propto \sqrt{\lambda} \phi_{0}^{3}   ~.\ee 
We will use these relations later when we consider the effects of curvature.

\section{Exact Analytical relations for the Q-ball chemical potential and energy per unit charge}

In this section we present a number of exact analytical relations between the Q-ball energy and charge for the case of a complex scalar with non-canonical kinetic terms. These results generalise the results for complex scalar field with canonical kinetic terms derived by Heeck et al in \cite{heeck21} (Eq.$(9)$ - Eq.$(12)$ of \cite{heeck21}). These relations are valuable as a way of confirming the stability of the Q-balls solutions without reliance on numerical solutions. This is particularly important for the case of the Q-balls arising from the non-minimal coupling to gravity, which have a new and more complex form of the Q-ball equation and for which the stability of the solutions is not intuitively apparent.

We present the details of the derivation of the results in Appendix A. The first relation is that between the change of the energy of the Q-ball and the change in its charge. In Appendix A we show that this is given by 

\be{ e107 } \frac{d E}{d \omega} = \omega \frac{d Q}{d \omega}  ~.\ee

\noindent This can also be written as

\be{ e108 }  \frac{d E}{ d Q} = \omega    ~.\ee 

\noindent Therefore $E$ and $Q$ will increase together, with the rate of increase given by $\omega$. This can assist in establishing which regions of the parameter space of Q-ball solutions correspond to stable and unstable Q-balls in a given model .

\noindent The second relation is that between energy per charge $E/Q$ and $\omega$. In Appendix A we show that this is given by   

\be{ e109 }   \frac{E}{Q} = \omega + \frac{4 \pi}{3 Q} \int dr \; r^{2} \frac{1}{\Omega^{2}} \left(\frac{d \phi}{d r} \right)^{2}   ~.\ee    

\noindent The integral depends on the gradient of $\phi$ and therefore can be interpreted as a surface energy term. This illustrates that, up to a correction due to surface energy, which we expect to be small (we will estimate the surface integral below using the analytic approximation), the $\omega$ parameter is equal to the energy per charge of the Q-ball. This is consistent with our results for $E/Q$ from the numerical Q-ball solutions. In the limit $\Omega \rightarrow 1$ this relation reduces to the corresponding relation derived in \cite{heeck21} for the case of conventional Q-balls with minimal coupling.  

Using the analytic approximation to estimate the surface integral, we find that (Appendix A)

\be{ e110 }    \frac{E}{Q} = \omega + \frac{\gamma}{30 \omega \phi_{0}^{2}}  ~\ee 
  
\noindent Since, for $m$ within the Q-ball window, 

\be{ e111  }   \gamma =  \frac{M_{Pl}^{2}}{\xi} \left(m^{2} + \omega^{2}  - \omega_{c}^{2} \right) \, \leq \frac{M_{Pl}^{2} m^{2}}{\xi}   ~\ee 

\noindent with equality when $m = \omega = \omega_{c}$, it follows that,

\be{ e112 }   \frac{E}{Q} \, \leq \, \omega  + \frac{\omega}{30} \frac{M_{Pl}^{2}}{\xi \phi_{0}^{2}} \left[1 + \frac{m^{2}}{\omega^{2}} \right]   ~\ee  

\noindent Therefore, since $\omega^{2} \thickapprox m^{2}$, we find that at most

\be{ e113 }  \frac{E}{Q} =  \omega 
\left[ 1  + O\left(\frac{M_{Pl}^{2}}{\xi \phi_{0}^{2}} \right) \right]       ~\ee 

\noindent Thus for Q-balls with $\phi_{0}$ well on the plateau of the potential, $\phi_{0} \gg M_{Pl}/\sqrt{\xi}$, we can say that $E/Q = \omega$ up to a very small correction.   This is consistent with the results obtained for numerical Q-ball solutions with $\phi_{0} \gg M_{Pl}/\sqrt{\xi}$. However, at small $\phi_{0}$, $E/Q$ can become significantly larger than $\omega$ as $\phi_{0}$ approaches the plateau limit. 

\section{Validity of the flat space approximation} 

In our analysis we have considered Q-balls in flat space. Here we consider the possible effects of curvature on our Q-ball solutions. We can consider the Q-ball solution to have a spherically symmetric energy density. The metric for this system is given by 
\be{k1} ds^2 = \left(1 - \frac{2 G M(r)}{r} \right) dt^{2} -  \left(1 - \frac{2 G M(r)}{r} \right)^{-1} dr^{2} - r^{2} d \Omega^{2}   ~,\ee
where $M(r)$ is the mass-energy contained within radius $r$. 
Therefore the effects of curvature will be negligible if 
\be{k3} R > 2 G M(R) \equiv r_{S} ~, \ee
where $R$ is the Q-ball radius, assuming that $M(r)$ increases more rapidly with $r$ than $r$ itself. In other words, the Q-ball should be sufficiently large compared to its Schwarzschild radius $r_{S}$. Comparing with the Q-ball radii and energies in Tables 1 and 2, and setting
$R = r_{Z}$ and $M = E_{Z}$, we find that this condition is easily satisfied for all of the Q-balls that we have numerically generated. The largest value of $r_{S}/r_{Z}$ is obtained for the largest value of $\phi_{0}$ in the tables, $\phi_{0} = 3.22 \times 10^{17} \GeV$, for which  $r_{Z} =  8.94 \times 10^{-10} \GeV^{-1}$,  $M = E_{Z} = 3.34 \times 10^{27} \GeV$, $r_{S} = 4.64 \times 10^{-11} \GeV^{-1}$, and $r_{S}/r_{Z} = 0.052$.

However, it is likely that Q-balls with larger values of $\phi_{0}$ will collapse into black holes once curvature is included. From our analytical approximation, we find that $E \propto \phi_{0}^{3}$ and $r_{Q} \propto \phi_{0}$. We also find that these relations are true for the values of $E_{Z}$ and $r_{Z}$ in the tables. 
Therefore we expect that 
$r_{S} \propto \phi_{0}^{3}$ and $r_{Z} \propto \phi_{0}$, and so $r_{S}/r_{Z} \propto \phi_{0}^{2}$. Normalising this relation using the Q-ball with $\phi_{0} = 3.22 \times 10^{17} \GeV$, we obtain the relation 
\be{k4}  \frac{r_{S}}{r_{Z}} = 0.052 \left(\frac{\phi_{0}}{3.22 \times 10^{17} \GeV} \right)^{2}   ~.\ee
This is a good fit to the numerical values for the Q-balls in the tables, particularly at large $\phi_{0}$. From this we expect that the value of $\phi_{0}$ at which the Q-balls will collapse to black holes once curvature is included, which we denote by $\phi_{0,\,c}$, corresponding to $r_{S}/r_{Z} = 1$, is 
\be{k5}  \phi_{0, \; c} = 1.41 \times 10^{18} \GeV  ~.\ee 
The mass and radius of this Q-ball are expected to be $r_{Z} = 3.91 \times 10^{-9} \GeV^{-1}$ and $E_{Z} = 2.80 \times 10^{29} \GeV$. (The values of $\phi_{0,\,c}$, $r_{Z}$ and $E_{Z}$ are expected to be somewhat less than these once curvature is fully included in the Q-ball solution, since as the black hole limit is approached we can expect gravity to reduce the radius of the Q-ball.) Therefore Q-balls with $\phi_{0}$ of the magnitude expected during tachyonic perturbation growth the end of Palatini inflation can directly collapse into black holes. We will discuss the possible implications of this in the next section.

The Q-balls considered above are for the case $\lambda = 0.1$. It is also interesting to determine how the value of $\phi_{0}$ and the mass of the smallest collapsing Q-ball vary with the self-coupling $\lambda$. From \eq{e1d} and \eq{e2d} we expect that for fixed $m/\omega_{c}$, the radius $r_{Z}$ will vary as $\sqrt{\lambda} \phi_{0}$ and the energy will vary as $E_{Z} \propto \sqrt{\lambda} \phi_{0}^{3}$. Using these, we find that 
$r_{S}/r_{Z}$ is independent of $\lambda$, and therefore the minimum value of $\phi_{0}$ for which a black hole forms is independent of $\lambda$,  whilst the mass of the black hole varies as $\sqrt{\lambda}$. 

In this preliminary analysis we have considered the flat space Q-ball solutions and determined the condition for these to form a black hole based on their associated Schwarzschild radius. In order to fully understand the properties of the Q-balls once curvature is included, an analysis of the complete curved space Q-ball solutions is required. Nevertheless, we expect that the flat space Q-balls will give a reasonable estimate of $\phi_{0,\,c}$ and so of the assocated Q-ball properties. To see this, note that the ratio of the Schwarzschild radius to the Q-ball radius, \eq{k4}, is proportional to $\phi_{0}^{2}$. Therefore the deviation of the metric \eq{k1} from the flat space metric at the surface of the Q-ball is proportional to $\phi_{0}^{2}$, e.g. for $\phi_{0} = \phi_{0,\,c}/2$ the deviation is by only 25$\%$. Moreover, this is the maximum deviation within the Q-ball, since we expect that $M(r) \propto r^3$ and so $M(r)/r \propto r^2$, therefore the metric will rapidly approach the flat space metric inside the Q-ball as $r$ decreases. Thus we do not expect curvature to significantly modify the flat space Q-ball solutions until $\phi_{0} > \phi_{0,\,c}/2$. Therefore the effect of curvature, although important, is not likely to very strongly alter the value of $\phi_{0,\,c}$ or the properties of the associated Q-balls from those of the flat space Q-ball estimates.   We note that the effect of curvature is likely to make the formation of black holes easier, as the additional gravitational attraction will tend to decrease the radius at which the attractive forces are balanced by the gradient pressure in the Q-ball solution for a given mass.

\section{Implications of Palatini Q-balls for Oscillons, Q-ball Theory and Cosmology.}

We have demonstrated the existence of Q-ball solutions of complex Palatini inflation for a range of the inflaton mass squared parameter. Moreover, we have shown that these Q-balls can have field values of  the expected magnitude of the inflaton field after slow-roll Palatini inflation and during the formation of non-linear lumps via tachyonic preheating, which are likely to be a precursor for the formation of non-topological solitons. In addition, we have shown that, for $\phi_{0} \gae 10^{18} \GeV$, the Q-balls will directly collapse into black holes once the effects of gravity are included.

In this section we will discuss the possible implications of our results for (A) the case of a real rather than complex inflaton field, (B) a new class of Q-balls with scalars non-minimally coupled to gravity, (C) inflaton condensate fragmentation, and (D) the post-inflation cosmology of Palatini inflation when non-topological soliton solutions exist. 

\subsection{Oscillons} 

One of our motivations for considering complex Palatini inflation is that it has exact solutions for the associated non-topological solitons (Q-balls) and so allows determination of the conditions for their existence. This is in contrast to the case of oscillons, where a non-trivial numerical solution is required. So understanding the properties of Q-balls and the conditions for their existence may indicate similar conditions for the existence of oscillons in the case of a real inflaton, which can serve as a guide to numerical searches for oscillon solutions in real Palatini inflation.  

The underlying dynamics of Q-balls and oscillons are similar. In both cases the existence of the objects can be understood as due to an attractive interaction between condensate scalars which is balanced by gradient pressure. The similarity in the dynamics is particularly evident in the I-ball approach to understanding oscillons, where the role of the conserved charge is replaced by a conserved adiabatic invariant \cite{kasuya03}. As such, we can conjecture that there will exist an analogous "oscillon window" in the inflaton mass squared similar to that for the Q-ball. 
Therefore a search for oscillons for values of the inflaton mass squared in the Q-ball window would, we believe, be likely to demonstrate the existence of Palatini oscillons which are consistent with inflation.

\subsection{A New Type of Q-ball} 

  The Q-balls we have discussed are, we believe, the first example of Q-balls with a non-canonical kinetic term for the complex scalar field. This results in a modified form of the equation for the Q-ball profile relative to the case of a conventional canonically normalised field. We have derived the analytical properties of the resulting Q-balls; in particular, we have shown that the relations between $E$, $Q$  and $\omega$, which are important for determining the stability of the Q-balls, remain either unchanged from the canonical case or are modified in a simple way. 

It is also interesting to consider the physical interpretation of the Q-balls in Palatini inflation. We have analysed the Q-balls in the Einstein frame, where the attractive interaction between the scalars is due to the plateau-type potential. Seen from the Jordan frame, the potential itself is actually repulsive in nature, being a combination of $\phi^2$ and $\phi^4$ terms. The attractive interaction is then entirely due to the non-minimal coupling of the scalars to the Ricci scalar, and is therefore entirely gravitational in origin and is sufficiently strong to overcome the repulsion of the $\phi^4$ potential term.   

\subsection{Inflaton Condensate Fragmentation} 

     In \cite{rubio19}, a linear perturbation theory analysis of the growth of inflaton perturbations immediately after the end of inflation was performed, which showed that sub-horizon inflaton perturbations rapidly grow to become non-linear. (We review this analysis in Appendix B.) This occurs for a range of perturbation wavelengths from the scale of the horizon to a minimum wavelength for which there is strong perturbation growth. Therefore in Palatini inflation it is expected that non-linear inflaton lumps will form at the end of inflation, with field magnitude of the order of the Planck scale. 

The question of the subsequent evolution of the non-linear lumps will then depend upon whether non-topological solutions exist. If such solutions exist, then it is likely that the non-linear lumps will evolve into non-topological solitons, in which case the inflaton condensate will fragment into non-topological solitons with vacuum between them. However, if such solutions do not exist, then since there is no solution for the field corresponding to stable objects within a vacuum, the non-linear lumps would have to evolve such that they are always embedded in a background field. In this case it is likely that the lumps will remain in contact with each other and will simply stretch and smooth out with expansion until the inflaton condensate returns to a homogeneous oscillating condensate. Therefore the existence or otherwise of non-topological solutions is crucial to the issue of fragmentation and the nature of the subsequent post-inflation era.  

\subsection{Post-Inflation Cosmology}

We next discuss the possible cosmological implications of complex Palatini inflation in the case where Q-balls (or their hypothetical oscillon counterparts) form at the end of inflation. 
This implies an alternative post-inflation cosmology compared to that of conventional Palatini inflation, which considers the inflaton mass to be effectively zero during inflation and reheating and therefore does not undergo inflaton condensate fragmentation.

\subsubsection{Non-topological Soliton-Dominated Era} 

In the case of a complex inflaton, we expect non-linear lumps of the real inflaton field along the radial direction will initially form. (At this stage we are not considering the possibility of a global $U(1)$ charge asymmetry being induced in the inflaton, as can happen in the case of non-minimally coupled inflaton Affleck-Dine baryogenesis models \cite{cline,cline2,adb1}.) Assuming that the existence of Q-balls implies the existence of related oscillon solutions, it is likely that an initial fragmentation to oscillons will then occur. In previous simulations of the evolution of real oscillons in models with a complex field (in particular, the numerical simulations of flat directions in gravity-mediated SUSY breaking in \cite{hiramatsu10}), it was found that such oscillons are unstable with respect to the formation of pairs of oppositely charged Q-balls\footnote{Even if oscillons do not form out of the non-linear lumps, it is conceivable that the non-linear lumps could themselves fragment into $\pm$Q-ball pairs.}. Therefore it is possible that, for inflaton masses squared within the Q-ball window, the inflaton condensate in complex Palatini inflation will evolve to a post-inflation era dominated by equal numbers of Q-balls with opposite charges. In this case the post-inflation era of complex Palatini inflation can become Q-ball-dominated and therefore quite different from conventional Palatini inflation. A similar oscillon-dominated post-inflation era may occur in the case of real Palatini inflation within an analogous oscillon window.

\subsubsection{Reheating via Non-Topological Soliton Decay}    

A non-topological soliton dominated era will have a impact on how reheating proceeds compared to the case of a homogeneous inflaton condensate. Reheating will occur via the decay and annihilation of the scalars in the solitons. In particular, reheating via Q-ball decay can proceed quite differently from a homogeneous condensate. Depending on the decay mode of the inflaton scalars, Q-ball decay can be slower than the decay rate of the individual scalars due to Pauli blocking, resulting in decay occurring only at the surface of the Q-ball
 \cite{cohen86}. In addition, annihilation processes between scalars that could occur for real inflatons will be excluded for the charged scalars forming the $\pm$Q-balls by global charge conservation. In particular, portal annihilation to the Higgs via $\lambda_{\Phi H}|\Phi|^{2} |H|^{2}$ would be excluded. An increase in the lifetime of the inflaton matter-dominated era due to Q-balls would decrease the number of e-foldings of inflation, modifying the predictions of inflation compared to the conventional case. On the other hand, portal annihilation processes in oscillons are actually enhanced relative the case of an inflaton condensate, as the annihilation rate, which is proportional to the number density squared, is constant within an oscillon, whereas the number density decreases as $a^{-3}$ (where $a$ is the scale factor) in a homogeneous inflaton condensate \cite{john02}.  

\subsubsection{Gravitational Waves and Primordial Black Holes from Non-Topological Solitons} 

The initial fragmentation of the condensate and formation of non-topological solitons could also have interesting consequences for gravitational wave and primordial black holes. Inflaton condensate fragmentation can result in production of gravitational waves \cite{gravfrag}. These are typically at very high frequencies (GHz), well beyond present gravitational waves detectors. However, there are on-going studies of possible ways to detect such gravitational waves, given their importance to early Universe cosmology \cite{ghz}. In addition, subsequent Q-ball decay can also produce characteristic gravitational waves \cite{km}. 

More specific to the possibility of formation of $\pm$Q-balls pairs, it has recently been proposed that fluctuations in the numbers of such Q-balls or oscillons formed at the end of inflation could lead to the formation of PBH \cite{kzpbh1}. (See also \cite{kzpbh2,kzpbh3,kzpbh4}.) In addition, in \cite{kzpbh5} it is suggested that PBH might be formed via mergers of Q-balls or oscillons. The model we have presented here could provide a well-motivated basis for such a Q-ball or oscillon model of PBH formation. 

In addition, we have found that, for sufficiently large field at condensate fragmentation, once the effect of curvature on the Q-ball solutions is included the corresponding Q-balls will either collapse to form black holes directly or could become massive enough to form black holes with only a few mergers.

We would also expect a distribution of Q-balls with different values of $\phi_{0}$ to result from tachyonic preheating and fragmentation. Depending on the details of the fragmentation process, the value of $\phi_{0}$ could either be large enough that almost all of the fragments form into black holes, or possibly only a few outliers with large $\phi_{0}$ will do so, with the majority of the energy density forming into Q-balls.

\subsubsection{Reheating and Poltergeist Gravitational Waves from the Decay of Black Holes due to Collapsed Non-Topological Solitons.}

The black holes resulting from gravitational collapse of Q-balls are not extremely heavy and will therefore decay during the early Universe. It is therefore possible that the Universe will become dominated by such black holes and that reheating will occur via their decay.
The evaporation time $t_{ev}$ of a black hole of mass $M \lesssim 10^{8} \, {\rm kg}$, for which the black hole temperature is greater than O(100) GeV, is \cite{evap}
\be{bh1} t_{ev} = \frac{40960 \pi}{27 g_{*\,BH}} G^{2} M^{3}  , \ee
where $g_{*\,BH} = 108.5$ is the number of spin degrees of freedom for the Standard Model. 
Equating this to the age for a matter-dominated Universe at reheating, $t = (2/3)H(T_{R})^{-1}$, where the expansion rate at reheating is $H(T_{R}) = k_{T_{R}} T_{R}^{2}/M_{Pl}$ (with $k_{T_{R}} \approx 3.3$), we obtain for the reheating temperature, 
\be{bh2}  T_{R} \approx \left(\frac{2}{3}\right)^{1/2} \left(\frac{27 g_{*\,BH}}{40960 \pi k_{T}}\right)^{1/2} \frac{8 \pi M_{Pl}^{5/2}}{M^{3/2}} =  1.145 \times 10^{6} \left(\frac{1 \, {\rm kg}}{M}\right)^{3/2} \GeV ~.\ee   
For the case of the Q-ball with the smallest $\phi_{0}$ for which a black hole will result once curvature is included, for which 
$\phi_{0} = 1.41 \times 10^{18} \GeV$ and  $M = E = 2.80\times 10^{29} \GeV (\equiv 498 \,{\rm kg})$ when $\lambda = 0.1$, we obtain 
\be{bh3}  T_{R} \approx 102.7 \GeV   ~.\ee
Therefore fragmentation to black holes in Palatini inflation, as suggested by the collapse of flat space Q-balls solutions once curvature is included, can naturally produce very low reheating temperatures. This can have significant consequences for freeze-out dark matter and other sub-TeV scale cosmology processes.

The mass of the smallest $\phi_{0}$ black hole is expected to be proportional to $\sqrt{\lambda}$, whilst the value of $\phi_{0}$ is independent of $\lambda$. Therefore we can reduce the value of $M$ and increase the reheating temperature by reducing $\lambda$. For example, with $\lambda = 0.01$ we expect that the smallest black hole will have $M = 158$ kg and the reheating temperature will be $T_{R} = 576.4 \GeV$. This may be advantageous for particle cosmology, as it raises the reheating temperature above the temperature of the electroweak phase transitiion \cite{rumm}, $T_{EW} = 159 \GeV$, and so allows for the possibility of baryogenesis via thermal leptogenesis or electroweak baryogenesis.

The very low reheating temperature has a strong effect on the prediction for the number of e-foldings $N_{*}$ corresponding to the Planck pivot scale $k = 0.05 \, {\rm Mpc^{-1}}$, and so for the predicted spectral index. For $T_{R} = 100 \GeV$ and $\lambda = 0.1$, and assuming complete matter domination by black holes from the end of slow-roll inflation (with initially $\rho_{bh} = \lambda M_{Pl}^{4}/4 \xi^{2}$), until black hole decay at $T_{R}$, we find that $N_{*} = 43.1$ (compared to $N_{*} = 51.9$ for instantaneous reheating at the end of slow-roll inflation). The resulting spectral index at the pivot scale is $n_{s} = 1 - 2/N_{*} = 0.9536$, which is excluded by the 2-$\sigma$ lower bound from Planck, $n_s = 0.9565$. (The range of values allowed by Planck is $n_{s} = 0.9649 \pm 0.0042$ (1-$\sigma$).) However, in a more complex reheating picture, where only a small fraction of the initial energy density is in the form of black holes from gravitational collapse of large Q-balls (which nevertheless come to dominate the energy density at late times) and the rest is in radiation from the decay of smaller Q-balls, the number of e-foldings of black hole matter domination will be smaller and so the value of $N_{*}$ will be larger. In this case $n_{s}$ could come within the Planck observational range\footnote{Alternatively, if the inflaton potential receives significant corrections, either quantum corrections due to the interaction of the inflaton with additional particles or non-renormalisable effective field theory corrections, and if these  raise the predicted value of $n_{s}$, then the reduction of $N_{*}$ and so of $n_{s}$ due to black hole matter domination can be consistent with the observationally allowed range of $n_{s}$.}. Thus the spectral index in this model will depend upon the details of the fragmentation process and the fraction of black holes formed. 

The possibility of the production of observable gravitational waves via the decay of primordial black holes by the Poltergeist mechanism was considered in \cite{inomata20}. In this mechanism, the rapid decay of PBH at the end of a PBH-dominated era causes sub-horizon density perturbations to become sub-horizon perturbations of the radiation density, which generate gravitational waves via sound waves in the thermal bath. It was found that the resulting gravitational waves could be detectable by future experiments (DECIGO and LISA) for black holes masses in the range 2-400 kg, although it should be noted that this requires that the black holes have a very narrow spread in mass, $\sigma\left(\delta M/M\right) \lae 0.01$. The upper bound of this mass range is close to the lower bound of approximately 500 kg on the range of black hole masses that we would expect to arise from Q-ball collapse at the end of fragmentation in Palatini inflation when $\lambda = 0.1$. We note that lower bound would be expected to be somewhat less than $500$ kg once the effects of curvature are included in the Q-ball solution, as the radius of the Q-ball for a given mass would be reduced by the gravitational attraction, leading to black hole formation at a lower mass. Also, smaller black hole masses are possible with smaller values of $\lambda$. Whilst it seems unlikely that the black holes from Q-ball collapse would have a sufficiently narrow spread of masses for observable Poltergeist gravitational waves to be generated, this can only be determined by numerical simulations.

Finally, we note that the existence of Q-balls that collapse into black holes once curvature is included raises the question of whether the non-linear fluctuations formed via tachyonic preheating could themselves collapse into black holes once curvature is included. This would impact Palatini inflation in general (including conventional Palatini Higgs Inflation), regardless of whether non-topological soliton solutions exist, and therefore motivates a study of tachyonic preheating including the effects of curvature.

\subsubsection{Dark matter from Inflaton Q-balls} 

We finally comment on the possibility of inflaton Q-balls playing the role of dark matter. Classically, Q-balls are absolutely stable. However, most of the inflaton energy density must decay to radiation to reheat the universe, making it difficult to understand how the inflaton Q-balls can remain stable or very long-lived. Nevertheless, if a small density in inflaton Q-balls can survive until the present then they could account for dark matter. This may be possible due to global charge conservation in the case where the inflaton condensate has a small global charge asymmetry. In this case one can consider the possibility that most of the energy density in the neutral non-linear lumps or initial oscillons can convert to radiation via $U(1)$-conserving annihilations, leaving only a very small density in Q-balls. If such Q-balls can survive in the radiation background, they could provide a dark matter candidate.

\section{Conclusions}  

   In this paper we have demonstrated the existence of Q-balls solutions of Palatini inflation with a complex inflaton. Palatini inflation is a favoured candidate for inflation, with excellent agreement with the observed spectral index and improved unitarity properties compared to conventional metric non-minimally coupled inflation. 

We have shown that Q-ball solutions exist for a range of inflaton masses, corresponding to a Q-ball window in the inflaton mass squared for which Q-balls can exist whilst being compatible with inflation. In determining this range we obtained a new upper bound on the possible value of the inflaton mass squared from the requirement that the potential can support inflation. The Q-ball window corresponds to inflaton masses squared close to the upper limit for which inflation is possible and represents 50$\%$ of the possible range of inflaton mass squared in terms of absolute value. The field inside the Q-ball can easily be of the order of the Planck scale, as is typical of the non-linear condensate lumps that form via tachyonic preheating at the end of Palatini inflation. 

Given the similarity of the physics of Q-balls and oscillons, we speculate that there may exist a similar oscillon window for the existence of oscillons for the case of Palatini inflation with a real inflaton. It would be interesting to numerically search for oscillon solutions for real inflaton masses within the Q-ball window. 

From a theoretical perspective, the Q-balls we have found constitute a new class of Q-balls, in which the complex field has a non-canonical kinetic term in the Einstein frame. This necessitates a modified analysis of Q-ball solutions as compared to conventional Q-balls based on canonical kinetic terms. In the Jordan frame, the attractive interaction responsible for holding the Q-balls together is entirely gravitational in nature, being due to the non-minimal coupling of the scalars to the Ricci scalar.

With respect to inflaton condensate fragmentation, we have emphasised that it is not sufficient for tachyonic preheating to occur in order for fragmentation to occur. Non-topological soliton solutions must also exist in order that the non-linear lumps from tachyonic preheating can evolve into fragments with vacuum between them. 

We have discussed the possible consequences of the existence of Q-balls (and their hypothetical associated oscillons) for Palatini inflation cosmology. These include modified reheating via Q-ball decay, gravitational wave production from fragmentation to Q-balls, primordial black holes from Q-ball number fluctuations or 
Q-ball mergers, and even dark matter due to stable Palatini inflaton Q-balls.

In addition, the flat space solution for Q-balls suggests that they form black holes if $\phi_{0} \gae 10^{18} \GeV$, which is typical of the inflaton field strength expected during tachyonic preheating at the end of Palatini inflation. Therefore it is possible that black holes may form either directly at condensate fragmentation or via merger of a small number of Q-balls. In addition, if the Universe becomes dominated by the black holes formed from the gravitational collapse of Q-balls, which have masses around 500 kg or more for the case with self-coupling $\lambda = 0.1$, then very low reheating temperatures, less than 100 GeV, will result. Smaller mass black holes and larger reheating temperatures are possible with smaller values of $\lambda$. A similar phenomenon may occur  in the case of a real inflaton via oscillons. The range of black holes masses expected from Q-ball collapse can be within the range 2-400 kg, for which observable gravitational waves can be produced via black hole decay and the Poltergeist mechanism. 
It would therefore be interesting to investigate the complete curved space Q-ball solutions for the case of large $\phi_{0}$, as well as the associated cosmology of fragmentation and reheating.  Moreover, this raises the question of whether the non-linear perturbations formed during tachyonic preheating in Palatini inflation could themselves collapse into black holes once curvature is included. This would impact Palatini inflation models in general, including conventional Palatini Higgs Inflation.    

Palatini inflation with a sufficiently large inflaton mass is likely to result in an alternative post-inflation cosmology that is quite different from conventional Palatini inflation. To begin to explore this possibility, a high-resolution numerical simulation of the tachyonic growth of perturbations and the subsequent fragmentation of the inflaton condensate, including the effects of curvature, will be necessary. We hope that this can be developed in the future.

\section*{Acknowledgements}

AKLS is supported by STFC.

 \section*{Note Added}   Recently a study of non-canonical Q-balls has been presented in \cite{lennon}. We find that our equation for the non-canonical Q-ball profile, \eq{e63},  agrees with the corresponding equation in \cite{lennon}.

\renewcommand{\theequation}{A-\arabic{equation}}
 \setcounter{equation}{0} 

\section*{Appendix A: Analytical Results for the Q-ball chemical potential and energy per unit charge}

In this section we derive exact relations between the energy and charge for Palatini non-minimally coupled Q-balls. This discussion follows very closely the derivation in Heeck et. al. \cite{heeck21} for the case of a minimally coupled complex scalar field (Eq.$(9)$ - Eq.$(12)$ of \cite{heeck21}).

\subsection{General relations between the Q-ball energy and charge} 

We start from the Q-ball energy functional \eq{e57}

\be{e116} 
E_{Q} = \int 4\pi r^{2} dr \left[\frac{1}{2\Omega^{2}}\left( \frac{\partial \phi}{\partial r}\right)^{2} 
 - \frac{\omega^{2}\phi^{2}}{2\Omega^{2}}
  + \frac{V}{\Omega^{4}}\right] + \omega Q.
\ee

\noindent Differentiating this with respect to $\omega$, and noting that $\phi \left(r \right) \rightarrow 0$ as $r \rightarrow \infty$ due to the Q-ball solutions matching with the $\phi =0$ vacuum at $r = \infty$, we obtain

\be{ e117 } 
\frac{d E}{d\omega} = \int 4\pi r^{2} dr \; \frac{\partial}{\partial \omega}\left[\frac{1}{2\Omega^{2}}\left( \frac{\partial \phi}{\partial r}\right)^{2}  - \frac{\omega^{2}\phi^{2}}{2\Omega^{2}}  
+ \frac{V}{\Omega^{4}}\right] + Q + \omega \frac{d Q}{d\omega},
\ee

\noindent where the Q-ball solution $\phi \left(r\right)$ for a given $\omega$ extremises $E_{Q}$  with respect to $Q$ to give the energy of the Q-ball, which we denote by $E$. At this point we use our earlier method of rescaling the field $\phi$ to $\sigma$, \eq{e69},

\be{e118} 
\frac{d E}{d\omega} = \int 4\pi r^{2} dr  \; \left[ \frac{1}{2}\frac{\partial}{\partial \omega}\left(\frac{\partial \sigma}{\partial r}\right)^{2} 
- \frac{\omega^{2}}{2}\frac{\partial}{\partial \omega}\left(\frac{\phi^{2}}{\Omega^{2}}\right) 
+ \frac{\partial}{\partial \omega}\left(\frac{V}{\Omega^{4}}\right)\right] + \omega \frac{d Q}{d\omega} ~.
\ee

\noindent The left hand side of the rescaled Q-ball equation in terms of $\sigma$, \eq{e72},

\be{ e119 }  \frac{d^{2} \sigma}{d r^{2}} + \frac{2}{r} \frac{d \sigma}{d r} = \frac{d V_{\omega}}{d \sigma} ~,\ee 
 
\noindent can be rewritten as

\be{ e120 } 
\frac{d^{2}\sigma}{d r^{2}} + \frac{2}{r}\frac{d \sigma}{d r} = \frac{1}{r^{2}}\frac{d}{dr}\left(r^{2}\frac{d\sigma}{dr}\right)
~,\ee

\noindent which can then be substituted back into the Q-ball equation and rearranged to give

\be{ e121 } 
\frac{\partial}{\partial \sigma}\left(\frac{V}{\Omega^{4}}\right) = \frac{1}{r^{2}}\frac{d}{dr}\left(r^{2}\frac{d\sigma}{dr}\right) + \frac{\partial}{\partial \sigma}\left(\frac{\omega^{2}\phi^{2}}{2\Omega^{2}}\right) ~.
\ee

\noindent We can write

\be{ e122  } 
\frac{\partial}{\partial \omega}\left(\frac{V}{\Omega^{4}}\right) = \frac{\partial}{\partial \sigma}\left(\frac{V}{\Omega^{4}}\right)\frac{\partial \sigma}{\partial \omega} ~.
\ee

\noindent Therefore

\be{ e123  } 
\frac{\partial}{\partial \omega}\left(\frac{V}{\Omega^{4}}\right) = \frac{1}{r^{2}}\frac{d}{dr}\left(r^{2}\frac{d\sigma}{dr}\right)\frac{\partial \sigma}{\partial \omega} + \frac{\partial}{\partial \sigma}\left(\frac{\omega^{2}\phi^{2}}{2\Omega^{2}}\right)\frac{\partial \sigma}{\partial \omega} ~.
\ee

\noindent Substituting this into \eq{e118} gives

\be{ e124  } 
\frac{d E}{d\omega} = \int 4\pi r^{2} dr \; \left[\frac{1}{2}\frac{\partial}{\partial \omega}\left(\frac{\partial \sigma}{\partial r}\right) \left(\frac{\partial \sigma}{\partial r}\right) -\frac{\omega^{2}}{2}\frac{\partial}{\partial \omega}\left(\frac{\phi^{2}}{\Omega^{2}}\right) + \frac{1}{r^{2}}\frac{d}{dr}\left(r^{2}\frac{d\sigma}{dr}\right)\frac{\partial \sigma}{\partial \omega} + \frac{\partial}{\partial \sigma}\left(\frac{\omega^{2}\phi^{2}}{2\Omega^{2}}\right)\frac{\partial \sigma}{\partial \omega}\right] + \omega \frac{d Q}{d\omega} ~.
\ee

\noindent  Rewriting

\be{ e125 } 
\frac{\omega^{2}}{2}\frac{\partial}{\partial \omega}\left(\frac{\phi^{2}}{\Omega^{2}}\right) = \frac{\omega^{2}}{2}\frac{\partial}{\partial \sigma}\left(\frac{\phi^{2}}{\Omega^{2}}\right) \frac{\partial \sigma}{\partial \omega}
\ee

\noindent gives

\be{ e126  } 
\frac{d E}{d\omega} = \int 4\pi r^{2} \, dr \, \left[\frac{1}{2}\frac{\partial}{\partial \omega}\left(\frac{\partial \sigma}{\partial r}\right) \left(\frac{\partial \sigma}{\partial r}\right) + \frac{1}{r^{2}}\frac{d}{dr}\left(r^{2}\frac{d\sigma}{dr}\right)\frac{\partial \sigma}{\partial \omega}\right] + \omega \frac{dQ}{d\omega} ~.
\ee

\noindent Integrating the second term by parts gives

\be{ e127 } 
\int \frac{1}{r^{2}}\frac{d}{dr}\left(r^{2}\frac{d\sigma}{dr}\right)\frac{\partial \sigma}{\partial \omega} dr = \left. \frac{\partial \sigma}{\partial r}\frac{\partial \sigma}{\partial \omega} \right|_{0}^{\infty}  - \int \frac{\partial \sigma}{\partial r} \frac{\partial^{2} \sigma}{\partial r \partial \omega}.
\ee

\noindent Due to the finite energy of the scalar field, the boundary term vanishes and we are left with

\be{ e128 } 
\frac{d E}{d\omega} = \int 4\pi r^{2} dr \left[\frac{1}{2}\frac{\partial}{\partial \omega}\left(\frac{\partial \sigma}{\partial r}\right) \left(\frac{\partial \sigma}{\partial r}\right) -  \frac{\partial \sigma}{\partial r} \frac{\partial^{2} \sigma}{\partial r \partial \omega} \right] + \omega \frac{dQ}{d\omega}.
\ee

\noindent The terms in the integrand cancel and we are left with 

\be{ e129  } 
\frac{d E}{d\omega} = \omega \frac{dQ}{d\omega}.
\ee

\noindent This is the same relation as that obtained for the case of minimally coupled scalars in \cite{heeck21}. This confirms that in the case of Q-balls in non-minimally coupled Palatini gravity, as in the case of a conventional minimally coupled complex scalar, $\omega$ acts as a chemical potential of the system, in that

\be{ e130 } 
\frac{d E}{dQ} = \omega.
\ee

\noindent This relation illustrates that $E\left(\omega \right)$ and $Q\left( \omega \right)$ will both increase and decrease together.

In addition to the chemical potential relation, it is possible to show analytically that the $\omega$ parameter is approximately equal to the ratio of the energy to the global charge, up to a small contribution from the surface energy of the Q-ball.  This derivation closely follows that of equation $(12)$ of Heeck et al. for minimally coupled scalars  \cite{heeck21}.

\noindent The energy of the Q-ball  is

\be{ e131 } 
E = 4\pi \int dr \, r^{2} \left[\frac{1}{2}\left(\frac{d \sigma}{d r}\right)^{2} + \frac{\omega^{2}\phi^{2}}{2\Omega^{2}} + \frac{V}{\Omega^{4}}\right].
\ee

\noindent The effective Q-ball action from \eq{e57} is 

\be{ e133 } 
\int dr \mathcal{L_{Q}} = \int 4\pi r^{2} dr \left[\frac{1}{2}\left(\frac{d\sigma}{dr}\right)^{2} -\frac{\omega^{2}\phi^{2}}{2\Omega^{2}} + \frac{V}{\Omega^{4}}\right] ~.
\ee

\noindent We make the rescaling $r \rightarrow \chi \tilde{r}$. In terms of the rescaled coordinate this becomes

\be{ e134 } 
\int d\tilde{r} \, \mathcal{L_{Q}} = \int d\tilde{r} \; 4\pi \tilde{r}^{2} \; \chi^{3} \left[\frac{1}{2\chi^{2}}\left(\frac{\partial \sigma}{\partial r}\right)^{2} - \frac{\omega^{2}\phi^{2}}{2\Omega^{2}} + \frac{V}{\Omega^{4}}\right]. ~
\ee

\noindent This expression carries both an explicit dependence on $\chi$ from the factors present in the integrand, and also an implicit dependence since $\sigma \left( r\right) \rightarrow \sigma \left(\chi \tilde{r} \right)$. Therefore a small variation in $\chi$ will cause a subsequent variation in $\sigma$. This means that

\be{ e135 } 
\frac{d\mathcal{L_{Q}}}{d\chi} = \left(\frac{d\mathcal{L_{Q}}}{d\chi}\right)_{explicit} +  \left(\frac{d\mathcal{L_{Q}}}{d\chi}\right)_{implicit}.
\ee

\noindent The Q-ball solution extremises the effective Q-ball action when $\chi = 1$, therefore $\int d\tilde{r}(d \mathcal{L_{Q}}/d \chi)_{implicit} = 0$  at $\chi = 1$. Since the integral is unchanged by rescaling the coordinates we require that

\be{ e136 } 
\int d\tilde{r} \left(\frac{d \mathcal{L}_{Q}}{d\chi}\right)_{explicit} = 0
\ee

\noindent when $\chi = 1$, where

\be{ e137 } 
\int d\tilde{r} \left(\frac{d \mathcal{L}_{Q}}{d\chi}\right)_{explicit} = 4\pi \int d\tilde{r} \; \frac{\partial}{\partial \chi}\left[ \frac{\chi}{2}\left(\frac{d\sigma}{dr}\right)^{2} - \frac{\chi^{3}\omega^{2}\phi^{2}}{2\Omega^{2}} + \frac{\chi^{3}V}{\Omega^{4}}\right]   = 
4\pi \int d\tilde{r} \; \tilde{r}^{2} \left[\frac{1}{2}\left(\frac{d\sigma}{d\tilde{r}}\right)^{2} + 3\chi^{2}\left(\frac{V}{\Omega^{4}} - \frac{\omega^{2}\phi^{2}}{2\Omega^{2}}\right)\right]      ~.\ee

\noindent Then, since $\tilde{r} \rightarrow r$ for $\chi = 1$, we have in this limit 

\be{ e138  } 
\int d\tilde{r} \left(\frac{d \mathcal{L}_{Q}}{d\chi}\right)_{explicit} = 4\pi \int dr r^{2} \left[\frac{1}{2}\left(\frac{d\sigma}{dr}\right)^{2} + 3\left(\frac{V}{\Omega^{4}} - \frac{\omega^{2} \phi^{2}}{2\Omega^{2}}\right)\right] = 0.
\ee

\noindent This can be rewritten as 

\be{ e139 } 
4\pi \int dr r^{2} \left[ \frac{1}{2}\left( \frac{d\sigma}{dr} \right)^{2} + \frac{3V}{\Omega^{4} } \right] = 4\pi \int r^{2} dr \frac{ 3\omega^{2} \phi^{2} }{ 2 \Omega^{2} } = 4\pi \cdot \frac{3}{2}\int r^{2} dr \frac{ \omega^{2} \phi^{2} }{ \Omega^{2} } = \frac{3}{2}\omega Q  ~.
\ee

\noindent Therefore

\be{e140} 
4\pi \int dr \; r^{2} \left[\frac{1}{2}\left(\frac{d\sigma}{dr}\right)^{2} + 3\left(\frac{V}{\Omega^{4} }\right) \right] = \frac{3}{2}\omega Q.
\ee

\noindent The Q-ball energy \eq{e118} can be written as

\be{ e141 } 
E = \omega Q + 4\pi \int dr \; r^{2} \left[\frac{1}{2}\left(\frac{d\sigma}{dr}\right)^{2} + \frac{V}{\Omega^{4}}\right] - \frac{1}{2} \omega Q ~,
\ee

\noindent where the last term is equivalent to

\be{ e142 } 
\frac{1}{2}\omega Q = 4\pi \int dr \; r^{2} \frac{\omega^{2}\phi^{2}}{2\Omega^{2}} ~.
\ee

\noindent From the relation \eq{e140} we have that

\be{ e143 } 
\frac{1}{2}\omega Q = 4\pi \int dr \; r^{2} \left[\frac{1}{6}\left(\frac{d\sigma}{dr}\right)^{2} + \frac{V}{\Omega^{4}}\right] ~. 
\ee

\noindent  Therefore

\be{ e144 } 
E = \omega Q + 4\pi \int dr \; r^{2} \left[\frac{1}{2}\left(\frac{d\sigma}{dr}\right)^{2} + \frac{V}{\Omega^{4}}\right] - 4\pi \int dr \; r^{2} \left[\frac{1}{6}\left(\frac{d\sigma}{dr}\right)^{2} + \frac{V}{\Omega^{4}}\right]  = \omega Q + \frac{4\pi}{3}\int dr \; r^{2} \left(\frac{d\sigma}{dr}\right)^{2} ~.
\ee

\noindent In terms of $\phi(r)$, the energy of the Q-ball is therefore

\be{ e145 } 
E = \omega Q + \frac{4\pi}{3}\int dr \; r^{2} \frac{1}{\Omega^{2}}\left(\frac{d\phi}{dr}\right)^{2} ~.
\ee

\noindent Therefore the exact expression for the energy per unit charge of the Q-ball is  

\be{ e146 } 
\frac{E}{Q} = \omega  + \frac{4\pi}{3Q}\int dr \; r^{2} \frac{1}{\Omega^{2}}\left(\frac{d\phi}{dr}\right)^{2} ~.
\ee

\noindent This illustrates that, up to a correction for the surface energy of the Q-ball that we expect to be small, the $\omega$ parameter is equal to the energy-charge ratio of the Q-ball. This also provides for an exact expression for the condition for absolute stability, $E < mQ$, in terms of $\omega$ 

\be{e147} 
\frac{E}{Q}  = \omega  + \frac{4\pi}{3Q}\int dr \; r^{2} \frac{1}{\Omega^{2}}\left(\frac{d\phi}{dr}\right)^{2} < m  ~.
\ee

\noindent If the surface term is significantly smaller than $\omega$ then the condition for absolute stability becomes $\omega < m$, which is the same condition as that for the existence of $Q$-ball solutions. Therefore in this limit all Q-ball solutions will be stable.

\subsection{Application to the analytical approximate Q-ball solution} 

We next use the analytical approximate solution to show that the energy to charge ratio is very close to $\omega$ for $\phi_{0} >> M_{pl}/\sqrt{\xi}$.
The analytical solution for $\phi(r)$ given by \eq{e104},
\be{ e148 }   \phi(r) = \phi_{0} \left[1 - \left( \frac{r}{r_{Q}} \right)^{2} \right]    ~, \ee 
is valid provided that the field does not deviate too far from its initial value $\phi \thickapprox \phi_{0}$, which is true for $r < r_{Q}/2$. Therefore the integral in in the expression for $E/Q$ is taken from 0 to $r = r_{Q}/2$.

\noindent Differentiating the field gives

\be{ e149  } 
\frac{d\phi}{d r} = -\frac{2r \phi_{0}}{r_{Q}^{2}} \Rightarrow \left(\frac{d\phi}{d r}\right)^{2} = \frac{4r^{2}\phi_{0}^{2}}{r_{Q}^{4}}.
\ee

\noindent By assumption the plateau approximation applies here, therefore

\be{ e150 } 
\Omega^{2} \thickapprox \frac{\xi \phi_{0}^{2}}{M_{pl}^{2}}.
\ee

\noindent Substituting these into \eq{e147} we obtain 

\be{ e151  } 
\frac{E}{Q} = \omega + \frac{4\pi \cdot 6\xi}{\pi \omega M_{pl}^{2}}\frac{1}{r_{Q}^{3}} \int^{r_{Q}/2}_{0} r^{2}\, dr \, \frac{M_{pl}^{2}}{3\xi \phi_{0}^{2}} \frac{4r^{2}\phi_{0}^{2}}{r_{Q}^{4}} = \omega + \frac{1}{5\omega}\frac{1}{r_{Q}^{2}} ~.
\ee

\noindent This can be written as

\be{ e152 } 
\frac{E}{Q} =  \omega + \frac{1}{30 \omega} \frac{\gamma}{\phi_{0}^{2}} ~, 
\ee

\noindent where

\be{ e153 }   \gamma = \frac{M_{Pl}^{2}}{\xi} \left(m^{2} + \omega^{2} - \omega_{c}^{2} \right)   ~.\ee

\noindent Since $m^{2} + \omega^{2} - \omega_{c}^{2} \leq m^{2}$ for $m$ within the Q-ball window, (with equality when $m = \omega = \omega_{c}$)  it follows that  

\be{ e154  } 
\frac{E}{Q} - \omega  \leq \frac{\omega}{30}\frac{M_{pl}^{2}}{\xi \phi_{0}^{2}}\left[ 1 + \frac{m^{2}}{\omega^{2}}\right]  ~.
\ee

\noindent  Therefore, since $\omega^{2} \thickapprox m^{2}$, we can say that, at most

\be{ e155 } 
\frac{E}{Q} = \omega \left[ 1 + \mathcal{O}\left( \frac{M_{pl}^{2}}{\xi \phi_{0}^{2}}\right) \right].
\ee

\noindent Thus we can say that $E/Q \thickapprox \omega$ for $\phi_{0} >> M_{pl}/\sqrt{\xi}$. Therefore in this limit we expect that the $\omega$ parameter in this model is equal to the energy-charge ratio to a very good approximation, and so $\omega < m$ will guarantee Q-ball stability.


\renewcommand{\theequation}{B-\arabic{equation}}
 \setcounter{equation}{0} 

\section*{Appendix B: Review of Tachyonic Preheating Results for Palatini Inflation}

We expect that the Q-balls in this model will arise as a result of the fragmentation of the inflationary condensate. The realisation of condensate fragmentation depends on a number of factors, and indeed can occur by different mechanisms depending on the inflationary model and the form of the inflaton potential \footnote{For a work which considers purely gravitational fragmentation for a general class of inflation models without self-resonance, see \cite{musoke19}}. 

In general, a 'flatter than $\phi^{2}$' inflaton potential will give an attractive interaction between the scalars, in turn creating a negative pressure within the inflationary condensate. This is a prerequisite for condensate fragmentation. Tachyonic preheating can be considered as a limiting case of perturbation growth due to negative pressure, in which the attractive interaction is so strong that scalar field perturbations grow more rapidly than the oscillation period of the field. There must also be a sufficient instability in the potential of the field such that the perturbations of the field can become non-linear in an expanding Universe. Finally, there must exist non-topological soliton solutions in order for discrete fragments to form 
\footnote{The process of perturbation growth and condensate fragmentation are very complex dynamically and the exact process is highly model dependent. This is a heavily simplified checklist to give a general idea of the more significant requirements that a model such as the one discussed in this work would need to produce Q-balls in practice.}\footnote{Other related phenomena such as Q-holes or Q-bulges can also form within a charged condensate in an appropriate potential without fragmentation occurring (see \cite{nugaev19}, \cite{nugaev16}).} \cite{coleman85}.

For the case where the inflaton in Palatini inflation is a real scalar field, it has been shown \cite{rubio19} that following from the onset of tachyonic preheating at the end of slow-roll inflation, there is a rapid growth of the inflaton fluctuations (within a single inflaton oscillation for non-minimal couplings of the size we are concerned with here). This allows the non-linear fluctuations to become the dominant contribution to the dynamics of the inflaton field, which can seed the formation of non-topological solitons from fragmentation. In the absence of non-topological soliton solutions, the tachyonic instability simply generates some large fluctuations within the condensate that will eventually decrease in amplitude as the universe expands and the post-inflationary evolution continues as usual.

A check on the viability of Q-ball formation in Palatini inflation is to consider the size of the largest non-linear perturbations in comparison to the size of Q-balls with field strength comparable to those of the non-linear perturbations. Here we estimate the size of the largest perturbation using the results of \cite{rubio19}, where a linear analysis of the growth of perturbations during the oscillations of the inflaton field at the end of Palatini inflation was performed.  An important difference between the model we consider and that considered in \cite{rubio19} is that in our model the potential at small field values (smaller than the inflationary plateau) is dominated by a $\phi^{2}$ term, whereas  in the conventional Palatini inflation model considered in \cite{rubio19} it is dominated by a $\phi^{4}$ term. This means that the tachyonic growth of perturbations (which we expect to take place during the inflationary plateau part of the field oscillations) is not smoothed out by a repulsive $\phi^{4}$ interaction as the field oscillates through lower values. Therefore we expect somewhat stronger growth of the perturbations in our model than in conventional Palatini inflation. We can therefore use the analysis in \cite{rubio19} to provide a lower estimate on the perturbation growth in our model and therefore on the range of perturbation wavelengths which undergo tachyonic growth and become non-linear.

At the end of slow roll inflation, the inflaton begins to undergo oscillations. The value of the inflaton field at the onset of these oscillations in Palatini inflation is:

\be{e156}
\phi_{end} = 2\sqrt{2} M_{pl} \sqrt{\beta} ~
\ee

\noindent where $\beta = 0.1-1$, which gives $\phi_{end} \sim (1-3)M_{pl}$ for the models we are interested in. The field when the non-linear lumps form will be less than this due to expansion during tachyonic preheating, therefore we can expect field values in the range $10^{17}-10^{18} \GeV$.  For our $\omega = 0.707155\omega_{c}$ Q-ball, the field at $r = 0$ is $\phi_{0} = 3.2217991 \times 10^{17} \GeV$ and its radius is $r_{X} = 8.89\times 10^{-10} \GeV^{-1}$,  where the scalar self-coupling is $\lambda = 0.1$. We will compare the radius of this Q-ball to the spectrum of perturbation wavelengths which undergo tachyonic growth in the analysis of \cite{rubio19} as an estimate as to whether these Q-balls are likely to form. 

The authors in \cite{rubio19} introduce a quantity defined as the perturbation wavenumber $\kappa$, which is a rescaling of the physical wavenumber $k$

\be{e157}
\kappa = \sqrt{\frac{\xi}{\lambda}}\frac{k}{M_{pl}} ~.
\ee

\noindent They determine a range of perturbation wavelengths $\kappa_{min} < \kappa < \kappa_{max}$ for which strong tachyonic growth of the perturbations occurs, where $\kappa_{max} \thickapprox 0.4a$, where $a$ is the scale factor,  and $\kappa_{min}$ corresponds to the size of the horizon.
These will correspond to a range of physical wavelengths $\lambda_{min} < \lambda < \lambda_{max}$, for which the perturbations become non-linear, where $\lambda_{min}$ is determined by $\kappa_{max}$ and $\lambda_{max} = 2 \pi H^{-1}$.

From \eq{e157}, the maximum physical wavenumber $k/a$ for perturbation growth is

\be{e158}  \frac{k_{max}}{a}  = \sqrt{\frac{\lambda}{\xi} } \frac{\kappa_{max}}{a} M_{Pl}  = \frac{2 \pi}{\lambda_{min}}   ~.\ee 

\noindent Therefore, with $\kappa_{max} = 0.4 a$, we obtain

\be{e159}  \lambda_{min}  = \frac{2\pi}{0.4}\sqrt{\frac{\xi}{\lambda}}M_{pl}^{-1} \equiv 15.7 \sqrt{\frac{\xi}{\lambda}}M_{pl}^{-1} ~.
\ee

\noindent With some rearrangement this gives

\be{e160}
\lambda_{min} = 6.5 \times 10^{-13}  \left(\frac{0.1}{\lambda}\right)^{1/2} \left(\frac{\xi}{10^{9}}\right)^{\frac{1}{2}} \GeV^{-1} ~ .
\ee

\noindent So for $\lambda = 0.1$ and $\xi = 1.2 \times 10^{9}$ we obtain $\lambda_{min} = 7.16 \times 10^{-13} \GeV^{-1}$.

\noindent The value of $H$ at the end of slow roll inflation is given by

\be{e161}
H^{2} = \frac{V}{3 M_{pl}^{2}}  ~,
\end{equation}

\noindent where the potential on the plateau is

\be{e162}
V = \frac{\lambda M_{pl}^{4}}{4 \xi^{2}}  ~. 
\end{equation}

\noindent Therefore

\be{e163}
H^{2} = \frac{\lambda}{12 \xi^{2}}M_{pl}^{2} ~,
\ee

\noindent and so 

\be{e164}
H^{-1} = \left(\frac{12}{\lambda}\right)^{\frac{1}{2}} \frac{\xi}{M_{pl}} = \left(4.56 \times 10^{-9} \right) \left(\frac{0.1}{\lambda}\right)^{1/2} \left(\frac{\xi}{10^{9}} \right) \GeV^{-1} ~ .
\ee

\noindent For $\lambda = 0.1$ and $\xi = 1.2\times 10^{9}$, this gives $H^{-1} = 5.47 \times 10^{-9} \GeV^{-1} $.

\noindent The maximum physical perturbation wavelength is then

\be{e165}
\lambda_{max} = 2\pi H^{-1} = \left(2.87 \times 10^{-8}\right) \left(\frac{0.1}{\lambda}\right)^{1/2}\left( \frac{\xi}{10^{9}}\right) \GeV^{-1}  ~,
\ee

\noindent which, for $\lambda = 0.1$ and $\xi = 1.2 \times 10^{9}$, gives $\lambda_{max}= 3.4 \times 10^{-8} \GeV^{-1}$.

So for $\lambda = 0.1$ and $\xi = 1.2 \times 10^{9}$, strong perturbation growth occurs for wavelengths $\lambda_{min} < \lambda  < \lambda_{max}
$, where $\lambda_{min} = 7.16 \times 10^{-13} \GeV^{-1} $ and $\lambda_{max} = 3.4 \times 10^{-8} \GeV^{-1}$. This range covers the radii of all of the Q-balls we have generated in this model. In particular, the radius $r_{X}$ of the Q-ball with $\phi_{0} = 3.2217991 \times 10^{17}$ GeV is well within the range for strong perturbation growth. This means that, assuming all other conditions for formation are met, the formation of  $\phi_{0} \sim 10^{17}-10^{18} \GeV$ Q-balls from non-linear lumps formed during tachyonic preheating at the end of Palatini inflation is physically feasible.


\end{document}